# Quantitative Dried-droplet Morphology and Image Analysis for Screening Adulteration in Milk

**Neha Gautam**[1,*], **Sumita Mondal**[1], **Debanjan Das**[1,*], and **Purbarun Dhar**[2,*]

[1] Centre of Excellence in Affordable Healthcare, Indian Institute of Technology Kharagpur, West Bengal – 721302, India

[2] Hydrodynamic and Thermal Multiphysics Lab (HTML), Department of Mechanical Engineering, Indian Institute of Technology Kharagpur, West Bengal – 721302, India

**Corresponding authors*: neha21gautam12@gmail.com ; debanjands@iitkgp.ac.in ; purbarun@mech.iitkgp.ac.in

## ABSTRACT

Adulteration of milk with water, urea, calcium compounds and starch is still a widespread food safety problem, especially in areas where there is no access to laboratory-based chemical testing, and is a global threat to human food safety and security, especially for children and the elderly. We present the development of a reagent free screening method, based on droplet-evaporative deposition method, optical microscopy, and quantitative image analysis, for consistent detection and classification of milk adulteration. Droplets of Single Toned (ST, 3% fat), and Double Toned (DT, 1.5% fat) milk samples, adulterated with water, urea, calcium, and starch, respectively, at different concentrations were tested. Deposition patterns were characterized by image processing using radial intensity profile descriptors (area under the curve, and edge decay slope) and gray-level co-occurrence matrix (GLCM) texture features (contrast, correlation, energy, homogeneity, and entropy). The descriptors exhibit consistent adulterant-specific trends: water and urea adulteration resulted in increasingly smooth, more homogeneous deposits (decreasing contrast, increasing homogeneity), while calcium and starch adulteration resulted in structurally rougher deposits (increasing contrast, decreasing homogeneity). Urea was further distinguished in the two groups by a significant increase in homogeneity and entropy collapse at higher concentrations, whereas calcium and starch were distinguished by diverging area under curve (AUC) trends. Milk type (ST vs. DT) was resolved by a combined multivariate signature of the descriptors at baseline. Our findings show that a simple, two-level feature-based framework – first resolving milk type, then adulterant family, then specific adulterant identity – can be realized entirely from optical microscopy data without additional chemical reagents.

***Keywords*:** Milk adulterants, droplet microfluidics, Area under curve (AUC), GLCM, droplet evaporation, drying pattern

## 1. INTRODUCTION

Milk occupies a unique position in the global food security scenario; as a primary source of nutrition for infants and vulnerable populations, yet one of the most adulterated commodities in the dairy supply chain[1,2]. Economic incentives to increase volume, or suppress quality deficiencies have resulted in widespread adulterations of raw and processed milk, with water[3][4], starch[5], urea[6] and calcium compounds, especially in unregulated or loosely regulated supply networks. These adulterants are cheap, easily available and difficult to detect by casual visual or organoleptic inspection, but produce measurable deviations from true fat content, protein to solid ratios, and overall nutritional value. The repercussions are not only restricted to economic deception, but regular intake of adulterated milk is linked to gastrointestinal, renal, and developmental health hazards, particularly in children and the aged[7][8–10]. Although conventional detection techniques, such as titrimetric assays[11], spectrophotometry[12][13][14,15], chromatography[16], electrochemical[17,18], and immunochemical[19,20] are very sensitive and accurate, they require trained personnel, laboratory infrastructure, and processing times unsuitable for on-site or point-of-need screening, especially in resource-poor environments like rural collection centers, or small-scale dairy cooperatives. The gap has created growing need for low-cost and rapidly field-deployable alternatives that can signal adulteration without the need for extensive sample preparation or specialized reagents. This has opened the door for physics-based and image-based diagnostic approaches to complement existing analytical chemistry techniques.

Milk is a multiphase colloidal system and not a simple liquid. Milk consists of fat globules dispersed in an aqueous serum phase; casein micelles stabilized by electrostatic and steric interactions, whey proteins, lactose and dissolved mineral salts[21]. The proportions of these constituents in milk vary for Single Toned (ST), Double Toned (DT) and Full Cream milk, due to the process of standardization of fat and solids-not-fat (SNF) during processing. The compositional heterogeneity determines the rheological behavior, surface tension, ionic strength, and solute mobility of the fluid, and all these directly impact the behavior of milk during evaporation[22][23][24]. Addition of an adulterant perturbs this colloidal equilibrium in a species-specific manner. Water dilutes the solute concentration and reduces the viscosity[25]. Urea alters the ionic strength and impairs the protein-protein interactions[26]. Calcium salts modify the cross-linking of the micelles and their precipitation behavior[27]. Starch adds a new macromolecular gelling agent with its own transport and aggregation kinetics[28]. Each adulterant interacts differently with the colloidal structure of milk, and the resulting suspension carries a compositional “fingerprint” that should, in principle, be recoverable, not only by direct chemical assay, but by the physical pattern left behind when the suspension is allowed to dry[29].

When a droplet of a colloidal fluid is deposited on a solid substrate and allowed to evaporate under ambient conditions, it undergoes a well characterized sequence of physical stages governed by contact line dynamics, evaporative flux and internal fluid transport[30]. In the classical pinned contact-line mode, the droplet edge stays fixed and the contact angle decreases as evaporation continues. The outward radial flow of fluid from the droplet interior toward the periphery compensates for the loss of solvent mass. The evaporation-driven transport is not uniform on the droplet surface: the flux is highest in the vicinity of the contact line due to the geometric singularity in the vapor diffusion at the wedge-shaped edge, a phenomenon first rigorously described in the seminal work on the diffusion-limited evaporation of sessile droplets[31]. The resulting non-uniform evaporative flux is the main driver of the internal flows that ultimately determine where dissolved and suspended solutes are deposited as the droplet dries, making the evaporation process itself a critical determinant of the final dried pattern. The external capillary flow produced by pinned-contact-line evaporation transports suspended particles, proteins, and salts to the edge of the droplet where they are concentrated and eventually deposited as a dense ring around the edge of the dried droplet - the famous coffee-ring effect[32].

Glass is a common model substrate in studies of droplet drying because it is smooth, chemically inert, optically transparent and non-absorbing, such that the droplet can dry completely through surface evaporation, without the confounding effects of fluid loss into the substrate itself [33]. This makes glass especially suitable for isolating the intrinsic compositional signature of a fluid: because there is no competing imbibition pathway, differences in the final dried pattern can be directly attributed to differences in viscosity, surface tension, ionic strength, and solute composition, rather than to variations in substrate wettability or pore structure. The well-defined, reproducible contact-line pinning habitually observed on glass also renders it a preferable substrate for quantitative optical microscopy, as the deposits are flat, adhere reliably to the slide surface, and can be imaged with controlled illumination without the scattering and surface roughness artifacts introduced by fibrous or porous media[34]. For these reasons, glass slides have been extensively used as a standard substrate in fundamental studies of coffee-ring formation, Marangoni-driven pattern transitions and biofluid drying. This provides a controlled platform for the composition-morphology relationship established in the previous section to be probed without further substrate-driven complexity.

Although qualitative morphological differences between dried deposits are often visually obvious under microscopic inspection, the identification of adulterants in a robust and reproducible way requires objective, and quantitative descriptors. Radial intensity profiling quantifies spatial distribution of optical density from droplet center to its periphery. Features such as area under the intensity curve (AUC) and edge decay slope describe ring sharpness, edge decay behaviour, and overall uniformity of deposition. Also, texture-based descriptors derived from the Gray-Level Co-occurrence Matrix (GLCM), such as contrast, correlation, energy, homogeneity, and entropy are able to describe the fine-scale spatial arrangement of pixel intensities within the deposit, and in turn reveal granularity, crack density and local heterogeneity that radial profiling alone cannot resolve. Together, these complementary feature families constitute a mapping of the physically dissimilar drying morphologies into an organized quantified feature space for statistical comparison and automated classification[35]. The extraction of radial and textural features from dried deposits naturally suggests a data driven classification framework that can map these quantitative descriptors to milk type and adulterant identity. Machine-learning classifiers are well suited to this task as they are able to learn nonlinear, high-dimensional relationships between combinations of features and sample class that would be difficult to specify through manually derived threshold rules, particularly in light of the compositional overlap and subtle gradations of morphology that can occur between types and concentrations of adulterants[36–38][39–41]. An additional benefit arises from a hierarchical classification strategy, which first distinguishes milk type and then identifies adulterant identity within each milk type, by breaking down a complex multi-class problem into more manageable sub-problems.

Although a body of work exists on the physics of droplet evaporation, coffee-ring formation, and image-based classification in model colloidal and biological fluids, the problem of milk adulteration has not been approached from a droplet-drying perspective. Furthermore, to date, there are no reports in the literature on the effect of water, urea, calcium and starch adulteration on the morphology of dried droplets of milk, although each adulterant should, on physical grounds, perturb viscosity, ionic strength and macromolecular composition in distinct ways. Similarly, application of quantitative image analysis in the form of radial intensity profiling or GLCM-based textural characterization on dried milk deposits for the detection of adulterants remains unexplored and the diagnostic potential of this approach remains completely lacking. Our study fills this gap by systematically characterizing the dried droplet morphology of Single Toned and Double Toned milk adulterated with water, urea, calcium, and starch on glass slide substrates, using radial intensity and GLCM texture features extracted from microscope images. We develop a hierarchical classification framework to distinguish both milk type and adulterant identity from these physically grounded image features – establishing, for the first time, a quantitative link between milk adulteration and dried droplet pattern formation.

## 2. MATERIALS AND METHODS

### 2.1 Experimental Methodology

For standard milk sample preparation in lab, Single Toned (ST, fat content ~ 3%) and Double Toned (DT, fat content ~ 1.5%) milk (all samples from a nationalized milk distribution company) were procured locally. Four different adulterants, namely filtered potable water, urea (from Sigma-Aldrich), starch (from Sigma-Aldrich), and calcium carbonate (food grade, from Urban Platter website) were used. Water dilution of 0%, 10%, 20%, 30%, 40%, and 50% were prepared. Since ST and DT milk already has lower fat % as compared to full cream milk (fat content ~ 8%), all adulterants were added in the pure ST and DT milk samples directly.

For urea adulteration, a 20% urea stock solution was first prepared in deionized (DI) water. Subsequently, urea concentrations of 0.07%, 0.5%, 1.0%, 5.0%, and 8.0% were prepared by mixing the stock solution with pure DT and ST milk, while maintaining the total sample volume at 1 mL. Likewise, starch solutions of concentrations of 0.1%, 0.5%, 1.0%, 2.0%, and 5.0% were prepared in DI water. Since starch is not readily soluble in water, a 10% stock solution of starch was prepared and heated at 80 °C under continuous stirring at 400 rpm for two hours to ensure uniform mixing. After the solution reached room temperature, the starch solution was mixed with pure ST and DT milk samples to achieve the desired adulterant concentrations. A 1% stock solution of calcium carbonate was prepared in DI water, and 0.05%, 0.1%, 0.25%, and 0.5% samples were then prepared by mixing the stock solution with ST and DT milk samples.

For microscopic imaging purpose, a Leica DM 3000 microscope is used and images were captured at 5X magnification. Image processing is performed using a MATLAB code, and image processing workflow is shown (supplementary file Fig. S1).

### 2.2 Radial Intensity Distribution Analysis

Initially, image analysis was used to plot the radial intensity distribution over the image. The radial intensity distribution analysis (RIDA) conveys details about the intensity fluctuations over the radial cross-section of the images. Here, each image was segmented into 200 bins, and the pixel intensity for each bin was quantified. The acquired radial intensity variation was depicted as a one-dimensional plot, with the radius value on the x-axis and the radial intensity distribution on the y-axis. The developed code automatically identified the droplet centroid during image processing, and the intensity measurement was limited to the deposit region.

The radial intensity distribution profile (RIDP) gives direct information on the internal redistribution mechanisms upon evaporation. Gradual intensity decay profiles indicate sustained colloidal redistribution and diffuse deposition behavior, while fast edge decay is suggestive of localized peripheral accumulation and compact redistribution transitions. Oscillatory or heterogeneous radial profiles indicate non-uniform transport behaviour induced by aggregation, gelation, crystallisation, or ionic reconstruction processes.

### 2.3 Area Under Curve (AUC)

RIDP show the deposition of the material from the centroid of the droplet towards the outer edge. However, the AUC is a unique quantitative measure that indicates the entire redistribution characteristics of the dried droplet. The AUC value displayed the redistribution characteristics in the droplet evaporation. A higher AUC value indicates a longer radial dispersion, better intensity retention, longer colloidal transport channels and dispersed deposition behaviour, which is typically associated with crystallization-assisted transport, ionic aggregation or persistent capillary redistribution. In contrast, low AUC values are characteristic of radial intensity decay, localized deposition, high edge density, and low transport dynamics. The suppression of transport due to gelation, localized aggregation, or a sudden change in deposition was reflected by a decrease in AUC.

### 2.4 Edge Decay Slope

To quantitatively characterize the distribution pattern around the droplet perimeter, decay slope calculation was performed on the radial intensity distribution profiles. The edge decay slope (EDS) gives the rate of intensity decay at the edge of the dried droplet and hence provides important information on the peripheral accumulation, the sharpness of redistribution and the edge localization behavior during evaporation. The outer region of evaporating colloidal droplets is particularly sensitive to the events occurring during evaporation, like capillary flows, colloidal accumulation, crystallization, gelation, and redistribution by desiccation. Thus, the structure and decay features of the radial intensity distribution around the droplet edge can serve well to identify the dominant transit and deposition processes of diverse adulterants. For calculation of EDS, $r \geq 0.7$ corresponds to 30% region from the outer normalized edge. The degradation behaviour in this region was determined using linear regression. A strong negative EDS indicates a sharp intensity drop at the edge, a concentrated peripheral localization, a steep redistribution changes and a highly structured deposition pattern. A modest negative slope, or a slope near zero, indicates diffuse redistribution, smooth peripheral transitions, weak edge localization, and slow deposition broadening.

### 2.5 GLCM and Granular Morphology analysis

Additionally, Gray Level Co-occurrence Matrix (GLCM) analysis was conducted on the individual images of the deposits. In GLCM analysis, several features of the image, including contrast, correlation, energy, homogeneity, and entropy, were retrieved to offer information regarding the presence of protein and various adulterants in milk samples. To further investigate the drying-induced aggregation behaviour, the granular area fraction was assessed by image segmentation. This statistic represents the percentage of the area of the droplet covered by separate granular deposits, and it is a direct measure of local crystallization and aggregate formation.

### 2.6 Statistical Analysis

All descriptors were extracted from one micrograph for individual milk type and for individual adulterant concentration. Since no independent droplet replicates were obtained, no inferential comparison between individual concentration levels was attempted, and no measure of between-droplet reproducibility is reported. Instead, the statistical treatment was applied to the three levels of the design where true replication is present: the concentration series (5-6 levels per adulterant), the matched pairing of toned (ST) and double-toned (DT) milk across the 20 common experimental conditions, and the pooled sample set (n=40) used for multivariate description. Spatial dispersion in the deposit was measured by subdividing each micrograph into sectors of concentric annuli and radial ones, and calculating the descriptors of the GLCM independently in each region of interest (ROI). Regions of less than 500 pixels were discarded. The resulting ROI ensemble was summarized by the mean and a percentile bootstrap 95% confidence interval of the mean (10,000 resamples). These intervals characterize the spatial heterogeneity of a single deposit, not estimates of experimental reproducibility.

Monotonic dose-response behaviour was evaluated using Spearman rank correlation (ρ) between adulterant concentration and each descriptor. As the concentration series span up to three orders of magnitude, ordinary least-squares regression was performed on $\log_{10}$-transformed concentration with the zero-adulterant control included using an offset of half the lowest non-zero level. Coefficients of determination ($R^2$) and p-values of regressions are provided. For those adulterants which showed an approximately linear response on the native concentration scale, an indicative limit of detection was calculated as LOD = 3s_res/|m|, where s_res is the residual standard deviation of the regression and m is the fitted slope. We used the Wilcoxon signed-rank test to compare toned and double toned milk across 20 matched adulterant-concentration conditions and report paired t-test and Cohen's d_z along with it. Intercorrelation between descriptors was evaluated by computing Pearson and Spearman coefficients on the pooled dataset. The standardized descriptors were analyzed using principal component analysis and the separability of the four adulterant classes was evaluated by linear discriminant analysis and random forest classification using leave-one-out cross-validation, against a chance baseline of 25%. All analyzes were performed in Python 3 (NumPy, SciPy, scikit-learn, scikit-image). Significance was taken at $\alpha = 0.05$.

# 3. RESULTS AND DISCUSSION

## 3.1 Drying Physics and Morphological Characterization of Pure Milk Droplets

Milk is a homogenous colloid suspension of fat globules, lactose, minerals, casein micelles and water. The evaporation of a sessile drop at room temperature begins with the formation of a pinned contact line at the edge. During drying, the internal flow generates both capillary transport and Marangoni convection. The solid particles out-migrate with components such as protein, fat and lactose which then collect near the margins. Nucleation and crystal formation thereafter take place, leading to the final dried pattern with the distinctive morphology of the droplet. All observed feature trends can be interpreted in the framework of evaporation-driven capillary flow in a pinned sessile droplet. As shown by Deegan et al. [42], a drying droplet with a pinned contact line has an outward capillary flow, as liquid evaporating from the edge is continuously replenished by liquid drawn from the interior, carrying dispersed solute toward the perimeter – the mechanism underlying ring-type deposition in all four descriptor sets used in this study (AUC, EDS, GLCM texture parameters). The local evaporative flux close to the contact line is given by Eq.1:

$$J(r) \propto \left(1 - \frac{r^2}{R^2}\right)^{-\lambda}, \lambda = \frac{\pi - 2\theta_c}{2\pi - 2\theta_c}, \quad (1)$$

where R is the radius of the droplet and $\theta c$ the contact angle. The physical origin of the peripheral accumulation quantified by AUC and EDS is the flux divergence at the edge. The Péclet number ( $Pe = \frac{vL}{D}$ ) defines the solutes that gather at the edge and those that stay evenly spread out, where D is the solute diffusion coefficient and is related to the hydrodynamic radius $r_h$ via the Stokes–Einstein relation [43](Eq.2).

$$D = \frac{K_b T}{6\pi\mu r_h} \quad (2)$$

In our study $r_h$ differs by roughly four orders of magnitude across urea (≈0.3 nm), hydrated $Ca^{2+}$ (≈0.3–0.4 nm)[44], casein micelles (≈50–150 nm)[45], and starch granules (≈1–15 µm)[46], each adulterant occupies a distinct transport regime, which is the physical basis for the adulterant-specific GLCM and granular percentage signatures described in details in section 3.7 and 3.8.

Water dilution introduces no new chemical species instead simply rescales the total solute concentration fed into the same evaporative flow field, mechanistically following Eq.3.

$$C(t) = C_0 \frac{V_0}{V_t} \quad (3)$$

A lower $C_0$ delays the onset of the jammed protein/lactose front at the contact line and produces a milder gradient with no triggering of any new nucleation/aggregation process consistent with the fact that water exhibits the mildest feature shifts of all four adulterants and, importantly, no entropy collapse, as no new mechanism for ordering is introduced. The effect of water dilution on the important physicochemical processes is the diluting of the components by addition of water, hence the reduction of viscosity and solids, which increases the evaporation rate and the outward capillary flow. The low solid content in the milk sample provides fewer impediments to the passage of particles leading to a thin deposition at the outer edge, a more prominent coffee ring and a sparse center. The complete drying mechanism and microscopic images of dried water diluted ST milk droplet are illustrated in Fig. (1) and additional water dilution images for both ST and DT milk sample are shown in supplementary file Fig. S2.

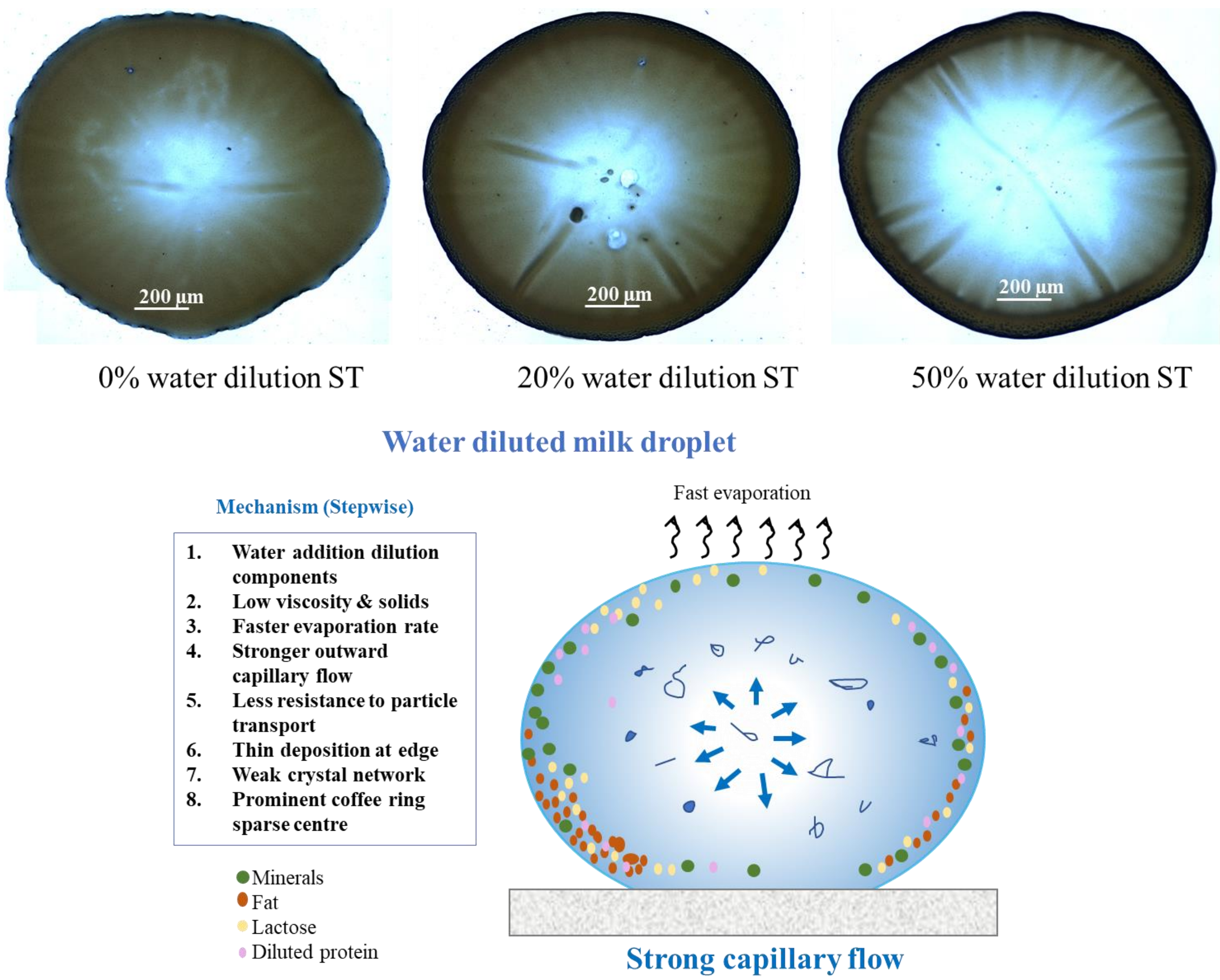

Fig. 1. Dried droplet morphologies of single-toned (ST) milk at different water-dilution levels (0–50%) and the proposed stepwise drying mechanism. Representative images demonstrate the evolution of the dried deposition pattern with increasing water dilution, while the schematic illustrates the sequential stages of evaporation, solute concentration, internal flow and particle redistribution, peripheral accumulation, and final residue formation.

Urea is a well-characterized chaotropic agent that perturbs the hydrogen-bonding network of water. By weakening the hydrophobic effect, it destabilizes protein native states. Molecular dynamics evidence supports a mechanism of direct interaction whereby urea forms stronger dispersion interactions with the protein backbone than does water, penetrating the hydrophobic core and breaking intra-backbone hydrogen bonds[47]. This predicts a progressive unraveling of the hydrophobic and calcium-phosphate-bridged network of the casein micelle at low to moderate urea concentration, which is consistent with the observed decrease in comparison. Second, and more importantly, urea has a finite solubility limit, so that as water evaporates the concentration increases according to Eq. 3. until C reaches $C_0$ at which point direct crystallization occurs.

$$CO(NH_2)_2(aq) \rightarrow CO(NH_2)_2(s) \tag{4}$$

Upon drying, it begins to exhibit needle-like crystals with a porous, uneven shape. The complete drying mechanism and microscopic images of dried water diluted ST milk droplet are illustrated in Fig. 2 and additional water dilution images for both ST and DT milk sample are shown in supplementary file Fig. S3.

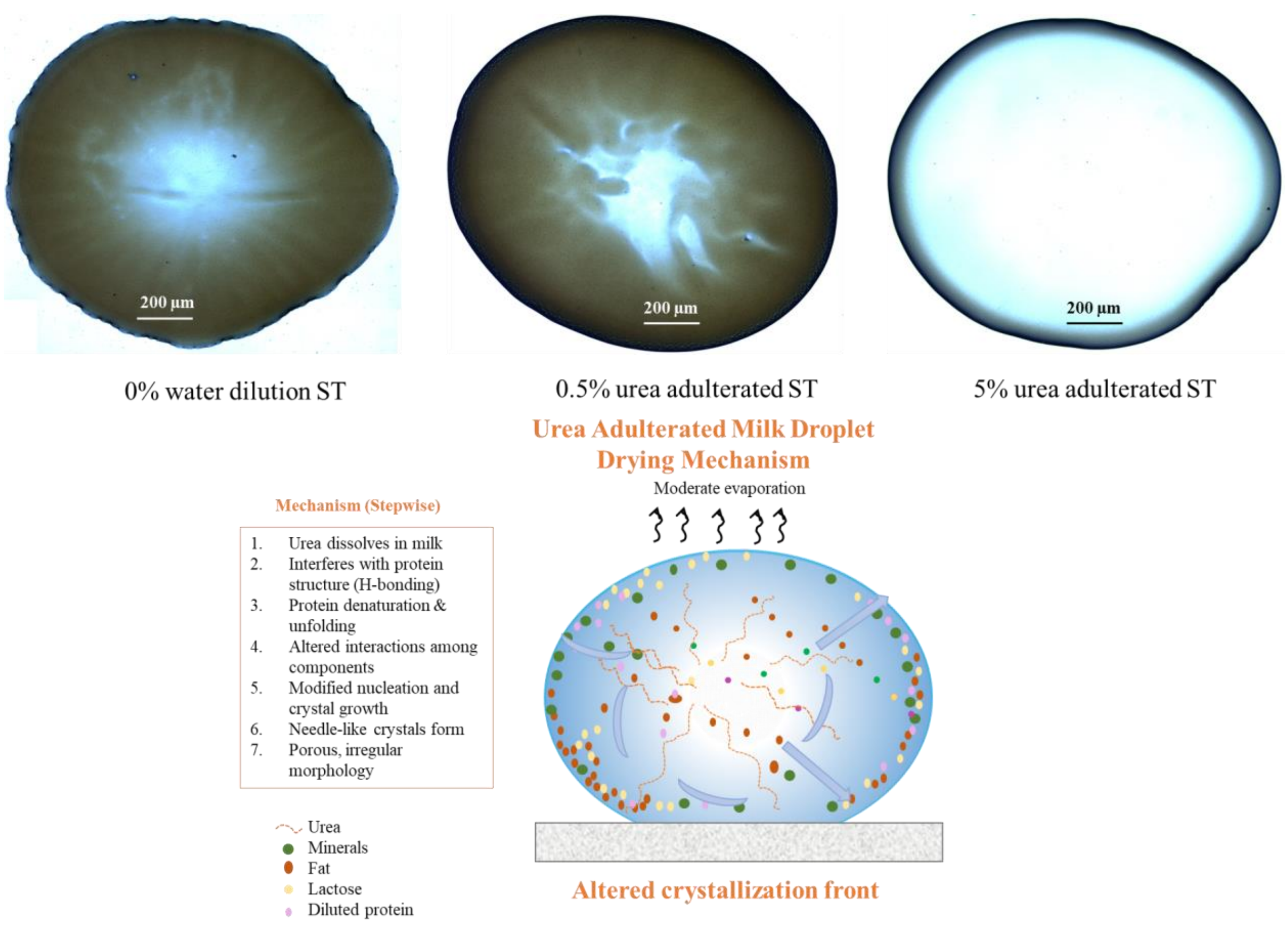

Fig. 2. Dried droplet morphologies of single-toned (ST) milk at different urea-adulteration levels (0–5%) and the proposed stepwise drying mechanism. Representative images demonstrate the evolution of the dried deposition pattern with increasing urea adulteration, while the schematic illustrates the sequential stages of evaporation, solute concentration, internal flow and particle redistribution, peripheral accumulation, and final residue formation.

Starch unlike the other adulterants, is purely physical and not chemical. Starch granules (≈2–30 µm, depending on source) are far larger than casein micelles or crystallizing urea/calcium species, and are effectively non-diffusive on the relevant drying timescale ($Pe \gg 1$) according to the Stokes–Einstein relation (Eq. 2). Importantly, the drying temperature used in this study (25 °C) is below the gelatinization onset of most common starches (≈60–70 °C), meaning that granules will not swell or form a gel network, but retain their structure. This intentional methodological choice is worth clarifying explicitly, as it confirms that starch behaves as an inert particulate tracer, rather than undergoing chemical transformation. The non-diffusive nature of starch granules means they are transported almost entirely by advection in the capillary flow, packing into a jammed heterogeneous layer at the drying front, directly accounting for the large increases in granular percentage and contrast observed.

The complete drying mechanism and microscopic images of dried water diluted ST milk droplet are illustrated in Fig. 3 and additional water dilution images for both ST and DT milk sample are shown in supplementary file Fig.S4.

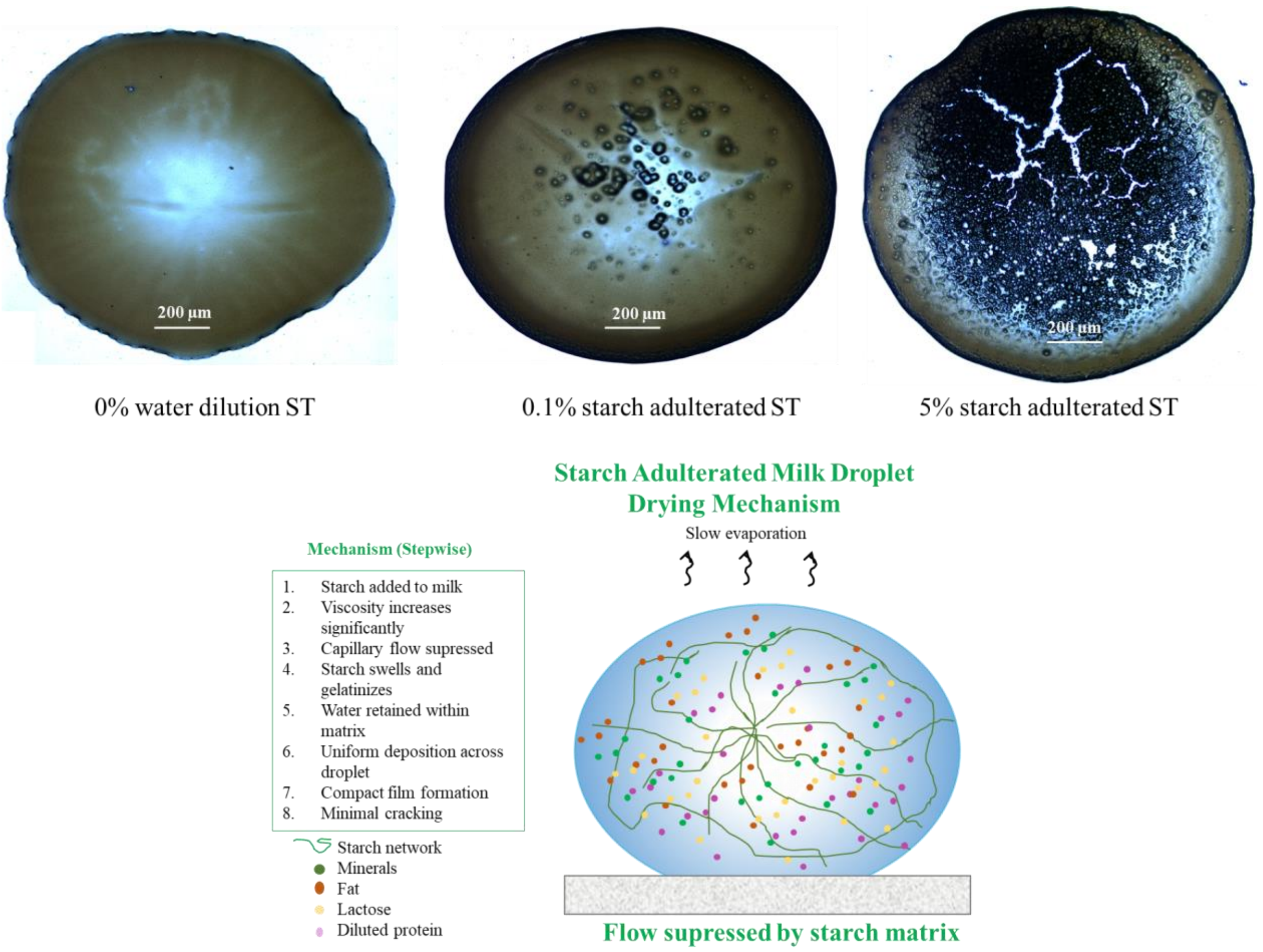

Fig. 3. Dried droplet morphologies of single-toned (ST) milk at different starch-adulteration levels (0–5%) and the proposed stepwise drying mechanism. Representative images demonstrate the evolution of the dried deposition pattern with increasing starch adulteration, while the schematic illustrates the sequential stages of evaporation, solute concentration, internal flow and particle redistribution, peripheral accumulation, and final residue formation.

In instances of calcium carbonate adulterated milk samples, calcium ions are introduced to neutralize the negative charges on casein micelles, thereby promoting micelle aggregation by diminishing electrostatic repulsion. In the DLVO model the electrostatic repulsion term decreases with increasing ionic strength (κ) and surface potential (ψ0) neutralization by bridging, decreasing inter-micellar repulsion and favouring partial coagulation [48]. The complete drying mechanism and microscopic images of dried water diluted ST milk droplet are illustrated in Fig. (4) and additional water dilution images for both ST and DT milk sample are shown in supplementary file Fig.S4.

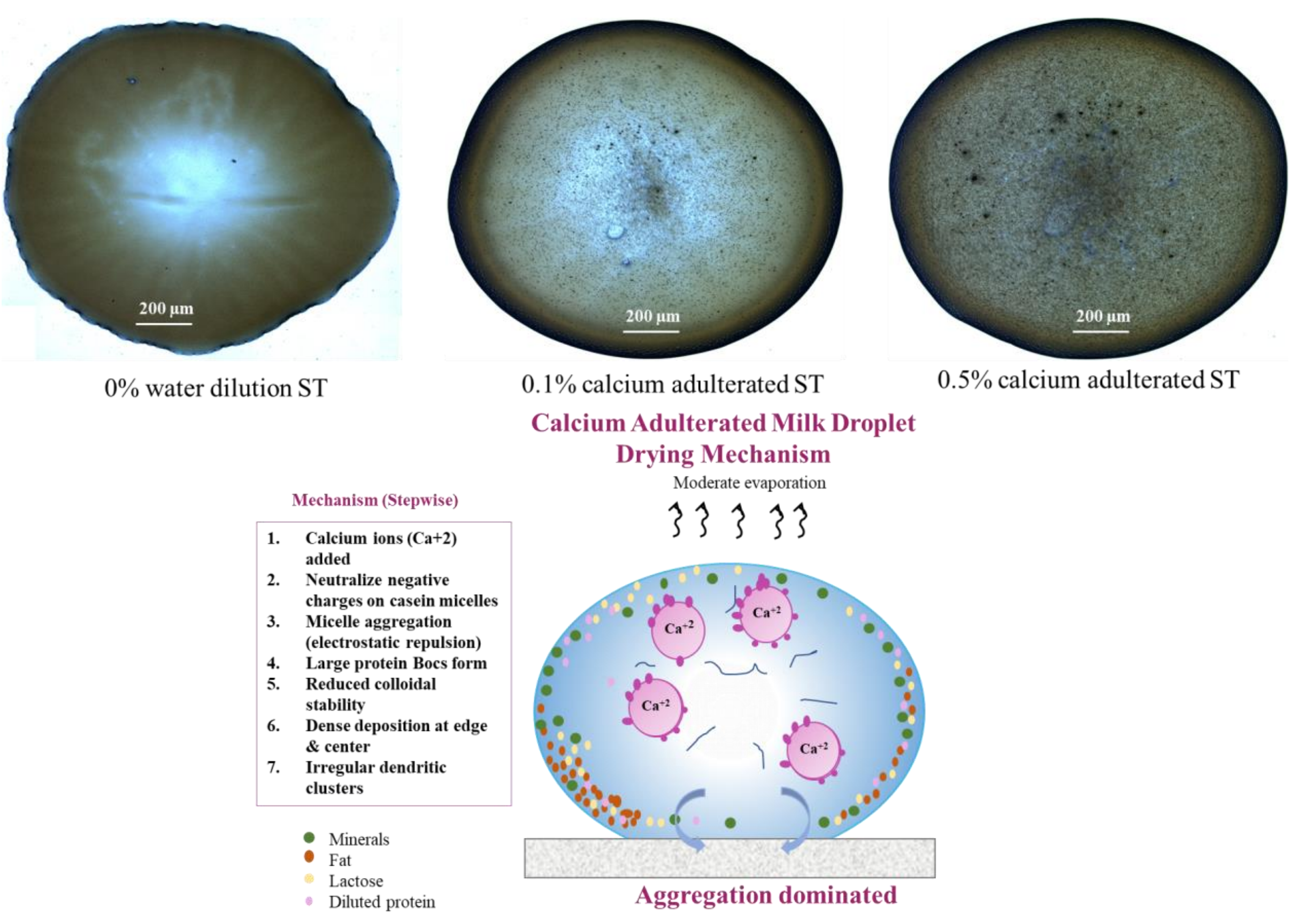

Fig. 4. Dried droplet morphologies of single-toned (ST) milk at different calcium-adulteration levels (0–0.5%) and the proposed stepwise drying mechanism. Representative images demonstrate the evolution of the dried deposition pattern with increasing calcium adulteration, while the schematic illustrates the sequential stages of evaporation, solute concentration, internal flow and particle redistribution, peripheral accumulation, and final residue formation.

### 3.2 Effect of water dilution on AUC and Radial Intensity Distribution

MATLAB image analysis was used to quantify the RID of the dried milk droplet patterns diluted with water. The comparison study of the radial intensity distribution showed that the diffusion zone was always larger in DT milk than in ST milk for water dilution because of its lower solid content and higher particle mobility during droplet evaporation.

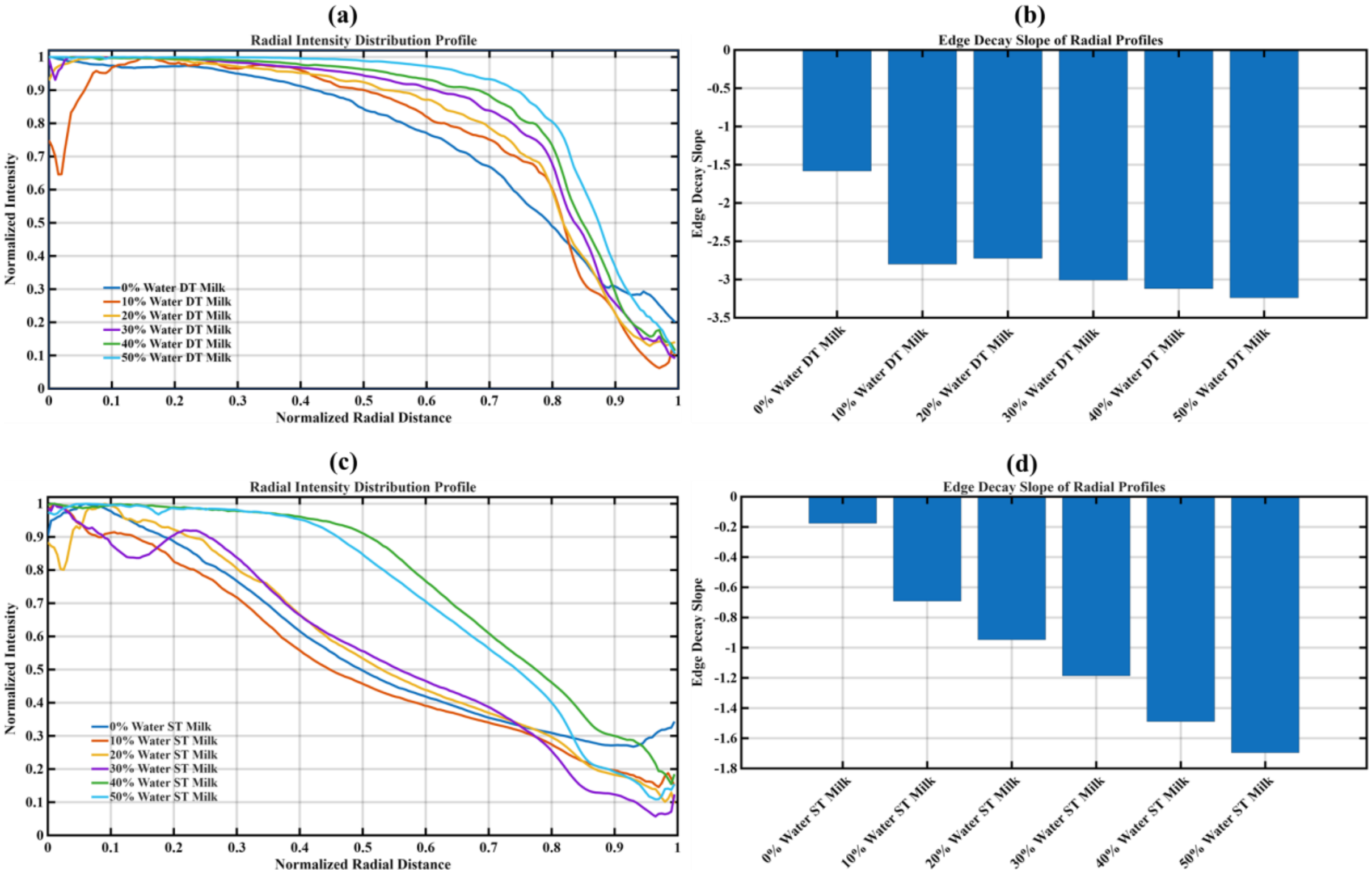

Fig. 5. RID and EDA of dried milk droplets at different water-dilution levels. Radial intensity profiles of dried (a) double-toned (DT) milk and (c) single-toned (ST) milk as a function of normalized radial distance, showing the spatial variation in normalized intensity from the droplet center toward the periphery. Corresponding edge decay slopes for DT milk (b) and ST milk (d) at different water-dilution levels are presented, quantifying the rate of intensity reduction near the droplet edge.

The obtained AUC value from the radial intensity profiles exhibited a consistent fluctuation with escalating water dilution. The quantitative image analysis for ST milk yields AUC values of 0.572, 0.523, 0.567, 0.556, 0.755, and 0.715 for 0%, 10%, 20%, 30%, 40%, and 50% water-diluted milk samples, respectively. The AUC values for DT milk samples are 0.748, 0.752, 0.785, 0.809, 0.829, and 0.863 for water dilutions of 0%, 10%, 20%, 30%, 40%, and 50%, respectively. Both the ST and DT milk samples show an increasing AUC value with the increasing water dilution level. This means that the increasing water level reduces the solid content and viscosity of the milk, which allows the liquid distributed in the capillary to stay for a longer time during the drying. Fig. 5 shows the pattern of radial intensity distribution of DT and ST milk samples. The values of EDS of water diluted milk samples are negative with values from 0.175 to 1.695 for the dilution percentage of 0%, 10%, 20%, 30%, 40%, and 50% accordingly for ST. The rising of EDS value for water dilution is observed at greater water dilution levels providing convincing evidence that higher water levels facilitate capillary-driven transport and significant diffusion of solid materials towards the outer periphery of the rim. Similar trend is observed for the DT milk sample with EDS value increasing from 1.582, 2.799, 2.723, 3.008, 3.119 and 3.239 at 0%, 10%, 20%, 30%, 40%, and 50% water dilution levels correspondingly. The DT milk sample had a higher raised EDS value because of the lower proportion of fat, which showed a more significant drop in the EDS for DT than that of ST milk. The EDS value for the water-adulterated milk sample is illustrated in Fig. 5.

### 3.3 Effect of urea adulteration on AUC and Radial Intensity Distribution

The incorporation of urea as an adulterant generates the most pronounced morphologically unique pattern of dried droplets in both ST and DT milk samples. Due to the inherently low-fat percentage of approximately 3.5% and 1.5% in both milk samples, the pure milk sample is augmented with varying concentrations of urea for the purpose of adulteration analysis. In ST milk, the concentration ring architecture was supplanted by irregular dendritic or starburst crystallization structures at urea concentrations of 0.5% or above, with substantial asymmetric aggregates dispersed across the centre zone. A nearly featureless or architecturally null deposit was detected in DT milk at higher concentrations (5-8%). The AUC increased considerably from 0.572 to 0.960 for 0 to 8% urea. The EDS is very uneven and non-monotonic (0.176-2.746). The non-monotonicity is mechanistically attributed to the chaotropic action of urea on the stability of the casein micelles, since urea perturbs the hydrogen-bonding network that stabilizes the quaternary structure of the casein and leads to partial protein unfolding. The unfolded chains aggregate via non-native hydrophobic interactions during drying, nucleating randomly instead of adhering to a structured outer boundary flow, resulting in the apparent dendritic architecture. Exceeding a catastrophic threshold (~5%) results in complete micelle disruption, thereby preventing coherent protein film formation and yielding a completely blank deposition signature in ST milk. The limit of detection (LOD) for urea is 0.40% (V/V) as determined by the AUC, being the lowest threshold among all adulterants examined in the current study. Consequently, this indicates the elevated sensitivity of the radial profile to even sub-percent concentrations of urea among all studied adulterants. For urea detection, the AUC is the chosen metric for quantifying indicators, while the morphological signature (dendrite or null deposit) offers independent visual confirmation. The quantitative image analysis of ST milk yields AUC values of 0.572,

0.555, 0.547, 0.648, 0.921, and 0.959 for urea adulteration levels of 0%, 0.07%, 0.5%, 1%, 5.0%, and 8%, respectively. The AUC values for DT milk samples are 0.746, 0.617, 0.685, 0.728, 0.897, and 0.957 for urea adulteration levels of 0%, 0.07%, 0.5%, 1%, 5.0%, and 8%, respectively. For both ST and DT milk samples, the AUC value increases with the rising urea concentration, indicating a persistent RID in the dried urea droplets. Fig. 6 illustrates the radial intensity distribution pattern for both DT and ST milk samples. The EDS values for urea-adulterated milk samples are negative, ranging from 0.810, 0.763, 1.213, 1.917, to 2.746 for 0.07%, 0.5%, 1%, 5.0%, and 8% urea-adulterated ST milk samples. A comparable trend is noted for the DT milk sample, with values rising from 1.7366, 1.8418, 1.2981, 2.9128, to 0.9218 for 0.07%, 0.5%, 1%, 5.0%, and 8% urea-adulterated DT milk samples. The EDS value for the urea-adulterated milk sample is illustrated in Fig. 6.

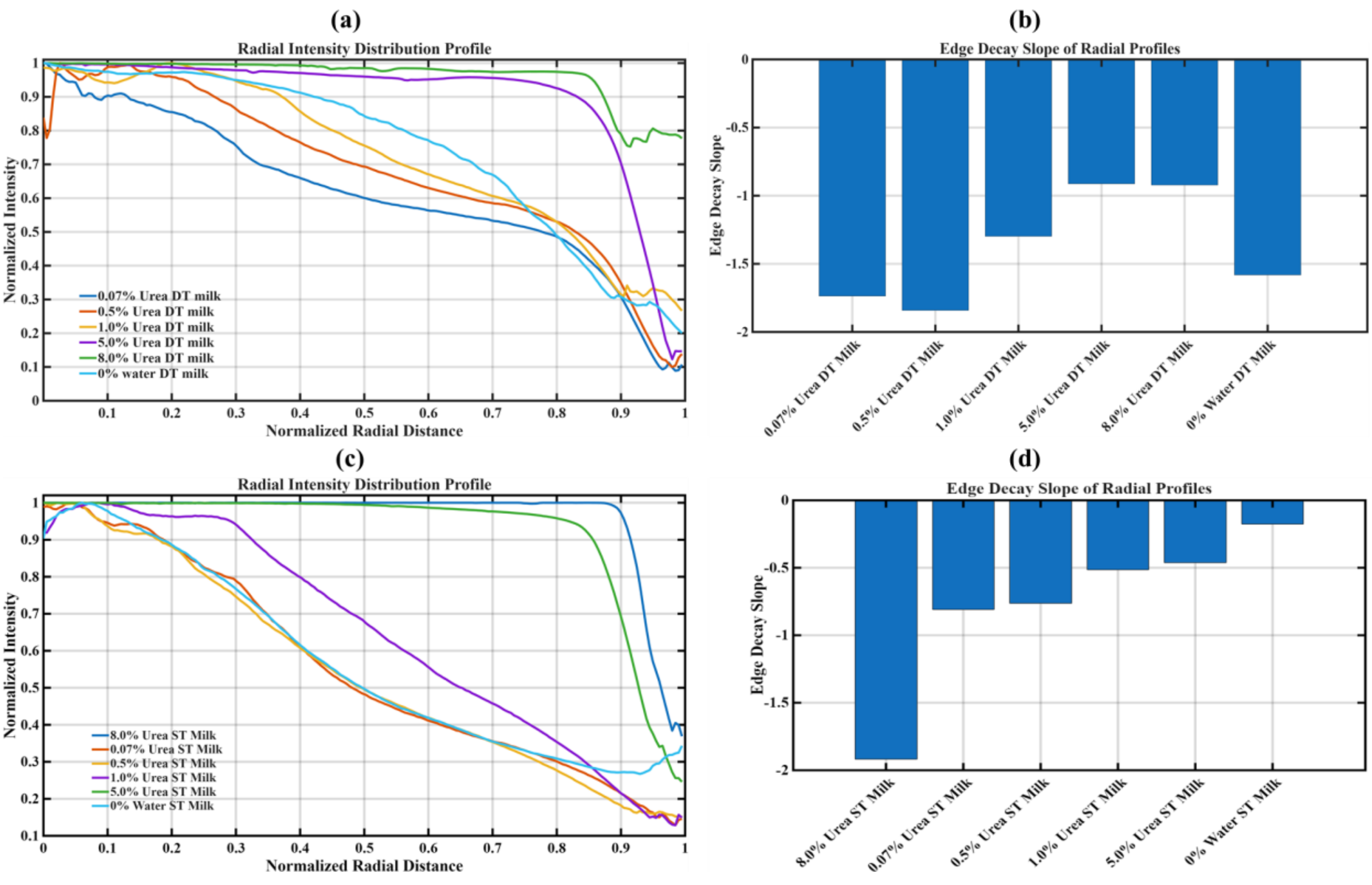

Fig. 6. Radial intensity distribution and edge decay analysis of dried milk droplets at different urea-adulteration levels. Radial intensity profiles of dried (a) double-toned (DT) milk and (c) single-toned (ST) milk as a function of normalized radial distance, showing the spatial variation in normalized intensity from the droplet center toward the periphery. Corresponding edge decay slopes for DT milk (b) and ST milk (d) at different urea-adulteration levels are presented, quantifying the rate of intensity reduction near the droplet edge.

### 3.5 Effect of Starch Adulteration on AUC and Radial Intensity Distribution

In our work, adulteration of starch produced the most optically distinctive and visible pattern, with high optical contrast at the particle level because of the inherent physicochemical incompatibility of the starch granules and the milk protein matrix. Profiles of starch-adulterated droplets exhibited a characteristic oscillatory profile with large irregular changes in intensity over the normalized radius (r/R = 0–1.0) not observed in the profiles of other adulterants. This oscillation arises from the non-uniform distribution of starch granules in the drying droplet. The preferential aggregation of the sedimented granules results in the development of alternating high and low optical density zones along the radial axis, interrupting the otherwise steady decrease in intensity from the centre to the periphery. AUC was low and non-monotonic quantitatively (0.534-0.672, ST: 0.58165-0.753, DT), and EDS varied between (-1.579 to -2.825, ST: -3.057 to -2.811, DT), showing the stochastic property of granule distribution. The AUC value decreases with starch adulteration for both ST and DT milk samples. The detection level (LOD) by the slope metric (5.27%) exceeds that of water or urea yet the oscillating profile pattern provides a unique visual and computational identifier which may be visually identified without quantitative calibration. The EDS value for the starch-adulterated milk sample is illustrated in Fig 7.

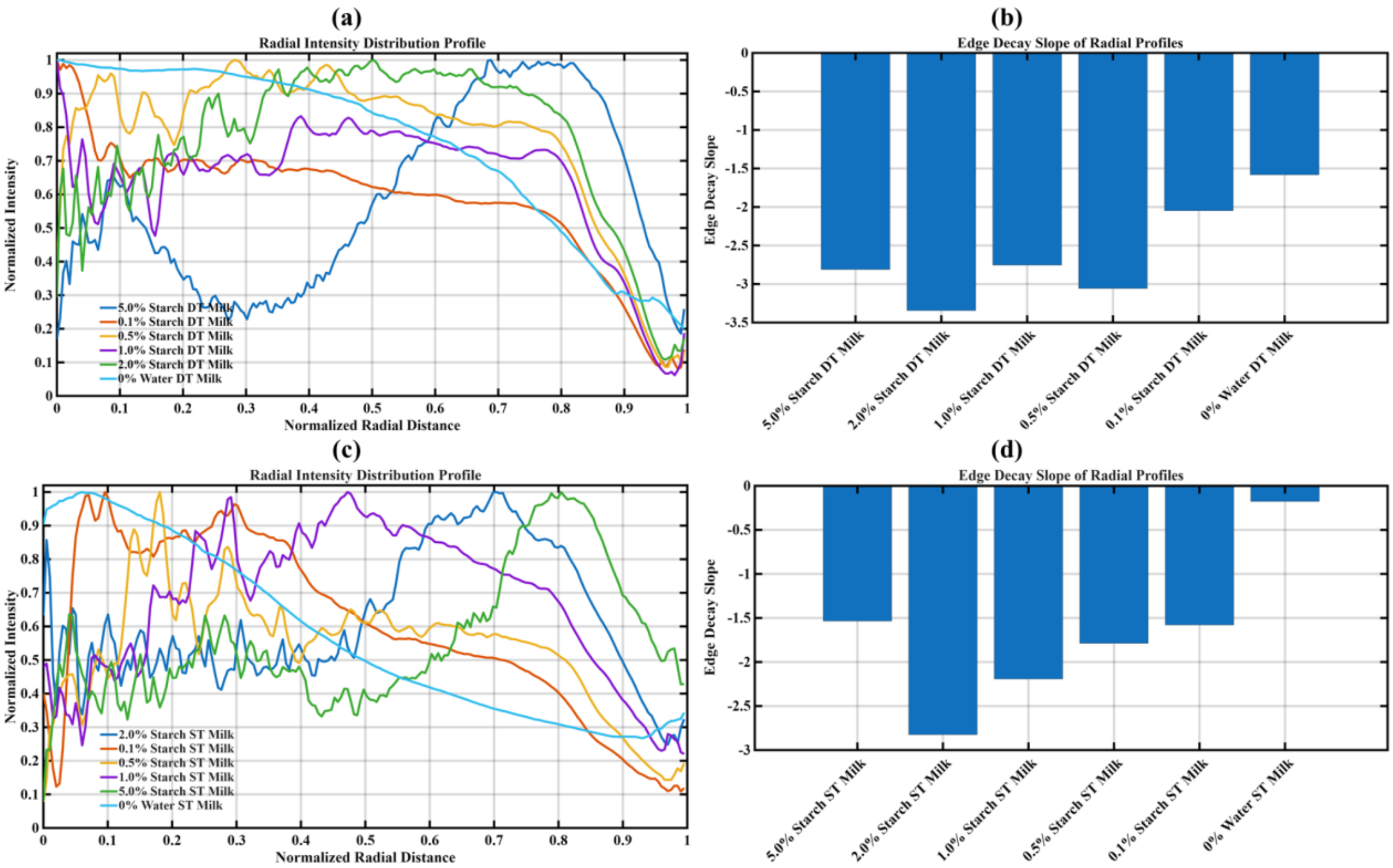

Fig. 7. RIDP and EDS analysis of dried milk droplets at different starch-adulteration levels. RIDP of dried (a) double-toned (DT) milk and (c) single-toned (ST) milk as a function of normalized radial distance, showing the spatial variation in normalized intensity from the droplet center toward the periphery. Corresponding EDS for DT milk (b) and ST milk (d) at different starch-adulteration levels are presented, quantifying the rate of intensity reduction near the droplet edge.

### 3.6 Effect of calcium carbonate adulteration on AUC and Radial Intensity Distribution

The calcium adulteration (0.05-0.5% w/v, added as $CaCl_2$) led to a morphologically subtle but quantitatively distinct pattern. The AUC was higher than that of the ST milk control even at the lowest concentration tested (0.716 at 0.05% vs. 0.571 at 0% Ca) but saturation was reached rapidly with little increase over 0.05-0.5% (0.716-0.830), indicating that the structural stabilization of the protein film by $Ca^{2+}$ reaches a plateau at low concentrations. The EDS in ST milk increased from 0.175 (control) to 2.298 at 0.1 % and remained high (2.298-2.825) at higher dosages. Similar behaviour is found for DT milk sample AUC value is practically constant fluctuating in the range of 0.783 to 0.792 for Ca concentration varies from 0.05% to 0.5% and EDS value is increasing from 1.786 to 2.126 in negative direction.

This plateau behavior illustrates the effect of $Ca^{2+}$: calcium ions crosslink phosphoryl residues on neighboring casein molecules in and between micelles, reinforcing the colloidal network and resulting in a more even, dome-shaped deposition pattern during drying, compared to the peripheral concentration in pure milk. The EDS value for the calcium-adulterated milk sample is illustrated in Fig 8.

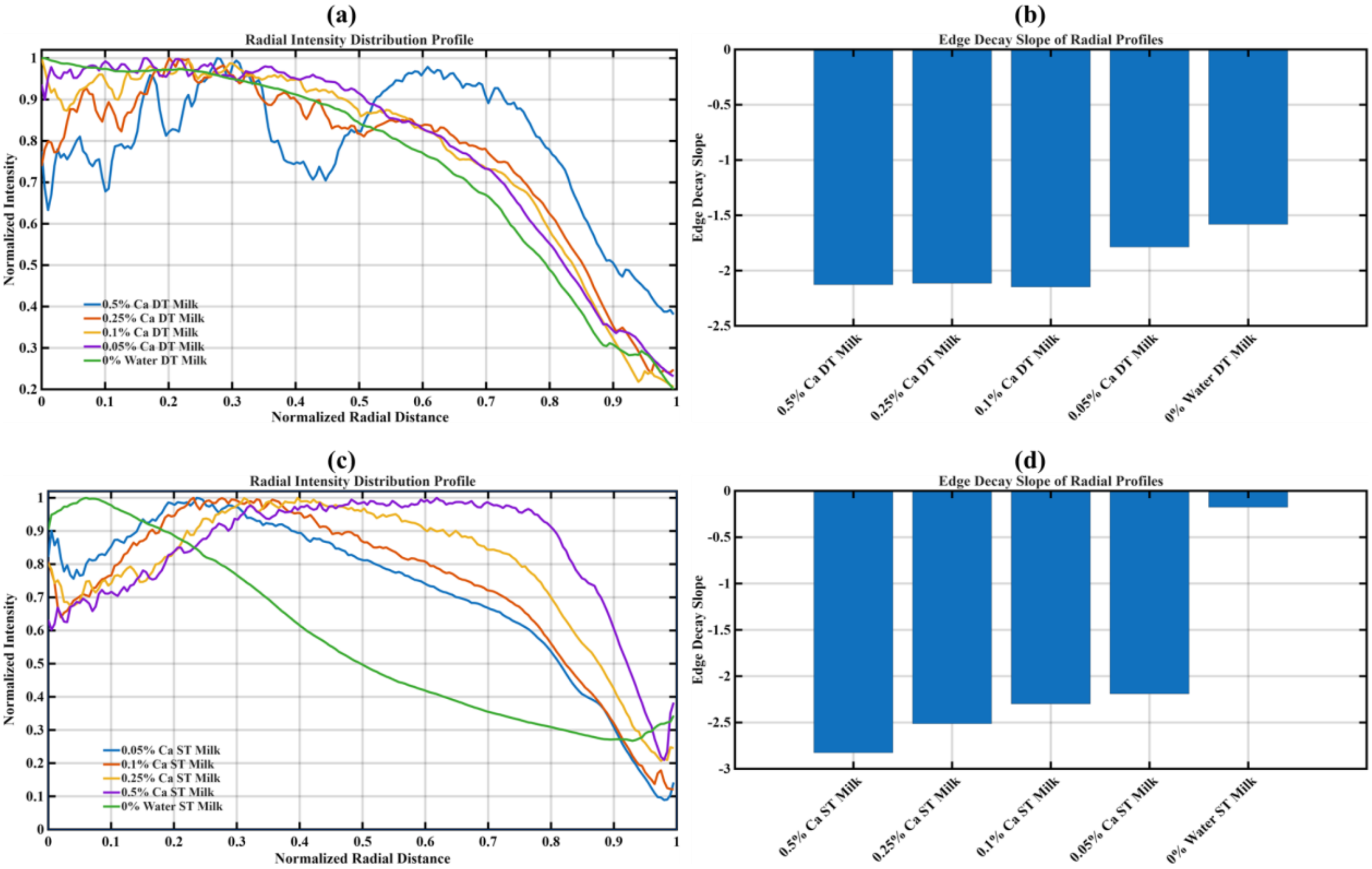

Fig. 8. RIDP and EDS analysis of dried milk droplets at different calcium-adulteration levels. Radial intensity profiles of dried (a) double-toned (DT) milk and (c) single-toned (ST) milk as a function of normalized radial distance, showing the spatial variation in normalized intensity from the droplet centre toward the periphery. Corresponding EDS for DT milk (b) and ST milk (d) at different calcium-adulteration levels are presented, quantifying the rate of intensity reduction near the droplet edge.

### 3.7 Texture Characterisation of Dried Milk Deposits Using GLCM Analysis

To further validate the microstructural alterations resulting from the incorporation of various adulterants in the milk sample, GLCM analysis was conducted on the individual images of the dried milk droplets. In GLCM analysis, several features of the picture, including contrast, correlation, energy, homogeneity, and entropy, were retrieved to offer information regarding the presence of protein and various adulterants in milk samples.

As the water dilution level in the ST milk sample increases, the contrast level gradually diminishes from 0.219 to 0.129, corresponding to a dilution range of 0% to 50%, attributable to the reduction in solid content resulting from the addition of water. Other characteristics such as correlation, energy, homogeneity, and entropy values stay constant across several water-diluted ST milk samples (Table S1**)**. The DT milk samples exhibit a similar pattern, with a decrease in contrast value from 0.1692 to 0.1211 while keeping nearly unchanged GLCM parameters (Table S2). This further substantiates that a similar trend is observed in both ST and DT water-diluted milk samples.

In instances of urea adulteration, urea is introduced at concentrations of 0.07%, 0.5%, 1%, 5%, and 8% in both ST and DT milk. The contrast value for urea in ST milk drastically decreases from 0.22234 to 0.0752 at elevated urea concentrations. A similar trend was observed with DT samples. The other parameters of GLCM show no significant changes. Urea changes the drying process dramatically by changing the hydration of proteins and the connections between protein molecules. At higher concentrations the disruption of the native casein network prevents production of irregular aggregates and enables formation of highly organized crystalline domains. The simultaneous increase of energy and uniformity, and decrease of entropy, implies the formation of a recurring and spatially structured microstructure.

The contrast value for ST milk dried droplets increased from a baseline value of 0.2197 to 0.3966 with the addition of adulteration of starch at concentrations of 0.1%, 0.5%, 1%, 2%, and 5%. The DT milk sample also shows a similar trend with starch levels increasing from 0.2143 to 0.4111. The increase in viscosity owing to starch limits the internal convective flow during evaporation, and hence prevents particle redistribution. Hence the starch rich regions are confined inside the droplet, leading to clustered polymeric aggregates and large intensity fluctuations over the dried deposit.

The contrast value for ST milk dried droplets increased from a baseline value of 0.2197 to 0.4951 with the addition of adulteration of calcium carbonate at concentrations of 0.05%, 0.1%, 0.025%, and 0.5%. The DT milk sample also shows a similar trend with calcium carbonate levels increasing from 0.1629 to 0.4457. The calcium ions increase the ionic strength of the milk sample enabling the localization of proteins and solid particles. The nucleation and crystal growth are enhanced, resulting in very heterogeneous drying patterns with large intensity differences, which increase contrast and entropy and reduce homogeneity.

Of all the GLCM measures obtained, contrast and homogeneity were the most vulnerable to adulteration. The correlation was always strong (0.96-0.99) and independent of the adulterant concentration, showing that the local spatial dependence between neighbouring pixels remained even when the texture shape changed. To summarize, water and urea created more uniform deposits with lower contrast and higher homogeneity, whereas calcium carbonate and starch enabled the formation of heterogeneous aggregates with significantly higher contrast. The results show that each adulterant affects a particular physicochemical process that governs the evaporation-induced self-assembly in the drying milk droplet.

**3.8 Quantitative Granular Morphology Analysis**

To further investigate the drying-induced aggregation behaviour the granular area fraction was assessed by picture segmentation. This statistic represents the percentage of the area of the droplet covered by separate granulary deposits, and it is a direct measure of local crystallization and aggregate formation. The granular area fraction decreased with increasing water concentration. The granular percentage in ST milk decreased from 1.05% in pure milk to 0.26% with 50% dilution while in DT milk it decreased from 0.09% to about 0.05%. Thus, the decrease in granular coverage shows that water dilution prevents the formation of localized aggregation by reducing the concentration of suspended particles. The reduced viscosity simultaneously increases the outward capillary transport resulting in a more homogeneous protein matrix with fewer discrete granular domains. The granular percentage was steady at low urea concentrations, but decreased dramatically at higher urea concentrations. Granular coverage in ST milk went down from ~1.0% to almost 0% and in DT milk from 0.77% to 0.03%. This behaviour shows that the formation of discrete aggregations is inhibited at high urea concentrations. The presence of urea favours the formation of continuous crystalline networks rather than discrete protein clusters, leading to a homogeneous distribution inside the droplet. This is in perfect agreement with the GLCM study, where increased energy and homogeneity are correlated with the disappearance of granular structures. Calcium carbonate increased the granular fraction in both milk samples: ST milk increased from 0.55% to 0.90% and DT milk from 0.21% to 0.51%. The increase in ionic strength favours heterogeneous nucleation, leading to crystalline aggregates occupying a bigger and larger fraction of the droplet area. Among all adulterants studied, starch showed the highest increase in granular coverage. The proportion of granules increased from 0.74 to 6.93 % in ST milk and from 0.97 to about 3.0 % in DT milk for starch adulteration. The increased viscosity due to starch impedes the mobility of particles within the evaporating droplet. Therefore, suspended starch granules and protein aggregates are spatially constrained and form enormous clustered regions, rather than redistributing towards the contact line. Thus, the greatly increased granular percentage is an indication of the increased incidence of polymer-mediated aggregation throughout the drying process.

The GLCM and granular analysis provide complementary information regarding drying behaviour of toned milk. Water dilution and urea adulteration simultaneously decreased granular coverage and GLCM contrast, and increased homogeneity, indicating smoother and more spatially uniform deposits. Conversely, calcium carbonate and starch adulteration increased both the granular percentage and the textural contrast, while reducing the homogeneity, which is consistent with the formation of highly heterogeneous crystalline and polymeric aggregates. The strong association between these several quantitative descriptors shows that the morphological changes are due to distinct changes in evaporation-driven colloidal transport, protein gelation and crystallization processes induced by each adulterant.

### 3.9 Identification of Milk Type and Adulterant by Two-phase Screening Framework

Taking the separate descriptors described in Sections 3.1–3.8, we investigated the potential to combine these traits into a hierarchical classification framework, capable of discriminating milk type, and subsequently identifying adulterant class and identification.

At baseline (0% adulterant) ST and DT milk were always separated for all four descriptor sets, with the highest relative difference being edge decay slope (EDS) (ST: −0.176 vs. DT: −1.582), followed by AUC (0.572 vs. 0.748) and GLCM homogeneity (0.924 vs. 0.959). This separation is consistent with the higher fat content of DT deposits resulting in measurably different drying-front dynamics.

However, this separation was not maintained in all adulteration series. At higher concentrations of urea (5-8%) and calcium (0.5%) ST and DT AUC and EDS values converged or crossed. This suggests that a single feature for milk-type classification is only reliable at low adulterant concentration and that a multivariate approach combining AUC, EDS, contrast and homogeneity is needed to maintain discriminability over the whole concentration range.

The GLCM contrast and homogeneity trends separate the four adulterants into two structurally different groups. Both water and urea adulteration correlated with decreasing contrast and increasing homogeneity relative to baseline, consistent with progressively smoother, more uniform deposit textures. A contrary trend was observed for calcium and starch adulteration with increasing contrast and decreasing homogeneity consistent with the introduction of structurally heterogeneous deposition features (e.g. crystalline or granular inclusions). This contrast/homogeneity directionality then acts as a coarse first-level discriminator between adulterant families independent of concentration.

Complementary descriptors could be used to further resolve the adulterant identity within each family. Water vs. urea: Urea-adulterated samples demonstrated a significant increase in homogeneity (approaching 0.99–0.998) and a corresponding decrease in entropy (as low as 0.37–1.43) at 5–8% concentration, consistent with increasingly ordered crystalline deposition. Water dilution did not reproduce this extremity at any concentration tested and granular percentage fell to exactly zero for urea at ≥5% but declined more gradually for water.

Calcium versus starch: Trends in AUC were divergent between these two adulterants – calcium AUC increased steadily with concentration in both types of milk while starch AUC remained relatively flat or slightly decreased. Moreover, the magnitude of the granular percentage was different, with starch having significantly higher absolute values (up to ~7% in ST milk) than calcium (~0.9%).

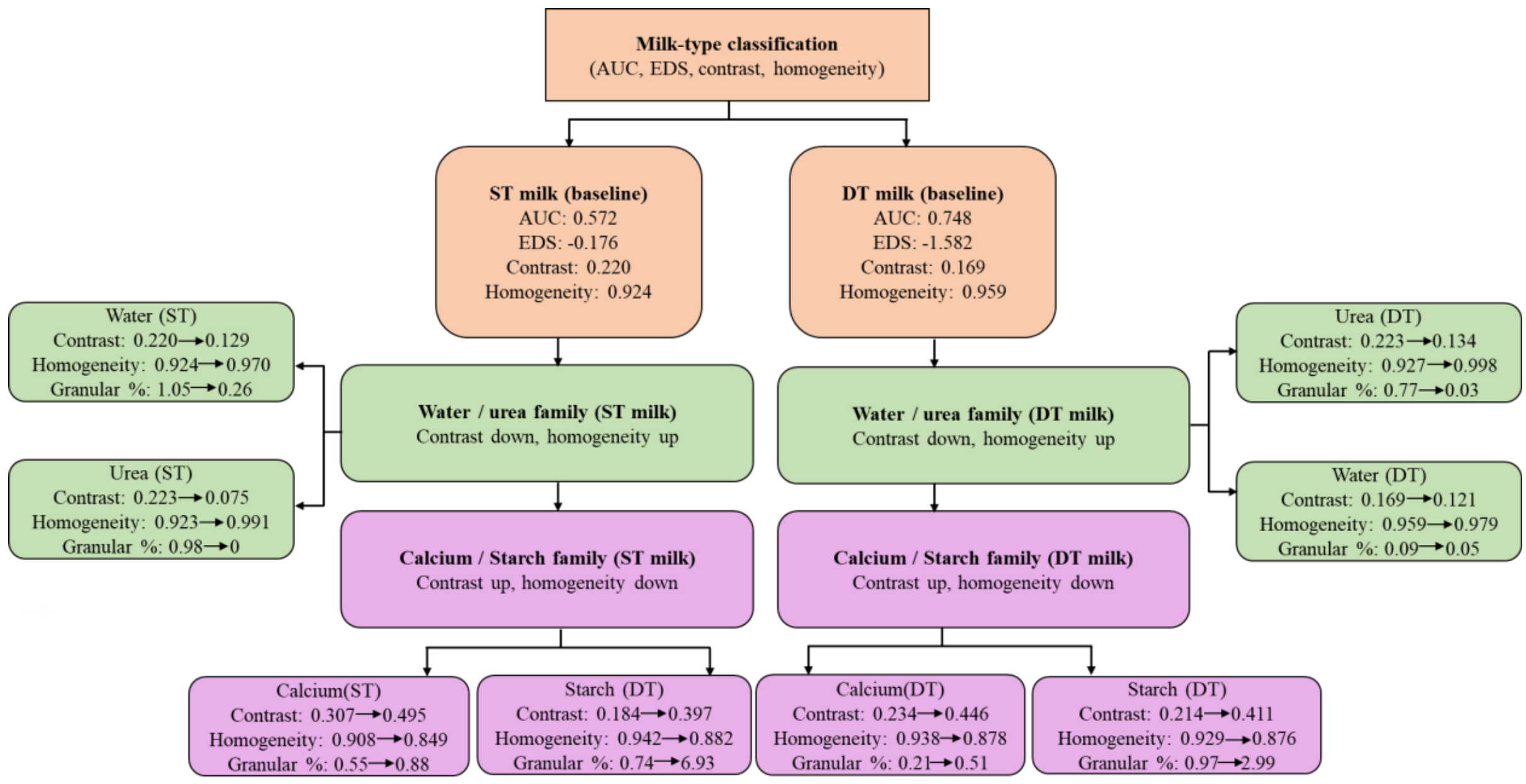

Fig. 9. Feature-based classification framework for distinguishing single-toned (ST) and double-toned (DT) milk and their adulteration patterns**.** The classification hierarchy first differentiates ST and DT milk based on radial intensity and texture descriptors, including AUC, edge decay slope (EDS), contrast, and homogeneity. Within each milk type, adulterants are further grouped into **water/urea** and **calcium/starch** families according to their characteristic changes in contrast and homogeneity. The corresponding changes in contrast, homogeneity, and granular fraction (%) illustrate the progressive differentiation of individual adulterants within each milk type.

### 3.8 Statistical characterisation of descriptor responses

Dose–response analysis revealed distinct trends specific to the adulterants in all three descriptors (contrast, homogeneity, granular area). Water dilution produced smooth and monotonic textural changes for both milk types, while calcium produced the strongest and most consistent contrast/homogeneity response, with detectability in the granular channel dependent on the background particulate loading of the milk. Starch had the greatest response of any adulterant in terms of granular area, consistent with its particulate nature, while urea exhibited a threshold type effect, with little change at low levels of adulteration and a sudden change at high levels of adulteration, most apparent in toned milk and to a lesser extent in double toned milk. The granular area was significantly different between toned and double-toned milk types, whereas contrast and homogeneity were similar across both indicating that textural descriptors are relatively robust to milk-fat variation while the particulate channel is more matrix-dependent.

Descriptor intercorrelation and PCA identified contrast and homogeneity as a single textural axis (strongly inversely correlated), with granular area as a largely independent particulate axis, explaining over 99% of total variance. Leave-one-out classification of the four adulterants (all three descriptors) achieved 65-72.5% accuracy (LDA and random forest, respectively) against an accuracy of 25% (chance level), with the main confusion between urea/water and starch/urea at low concentrations, consistent with their shared mechanism of reducing textural heterogeneity.

From indicative detection limits, calcium was the most sensitively detected adulterant (~0.1–0.3%), starch was detectable via the granular channel (~1.1%), and urea required higher concentrations for reliable detection, indicative of its threshold-driven response. Full statistical details (rank-correlation coefficients, regression outputs and detection-limit calculations) are given in the Supplementary Information section S1.

## 4. CONCLUSION

The study shows the utility of combining dried droplet evaporative deposition with optical microscopy and quantitative image analysis to discriminate milk types and identify several categories of adulterants without the need for chemical reagents. Distinct deposition signatures were observed in the radial intensity profile descriptors (AUC, edge decay slope) and GLCM texture parameters (contrast, correlation, energy, homogeneity, entropy) in all adulterated samples (water, urea, calcium, starch) of ST, and DT milk samples. The key contribution of this work is a two-tier classification logic that replicates the way a field deployable tool would have to operate. In the first tier, milk type was resolved using a combined signature of AUC, edge decay slope, contrast and homogeneity. However, this separation was strongest at low adulterant concentration and a multivariate combination rather than any single descriptor was required to remain robust at higher concentrations. The adulterant family was clearly separated at the second level by the direction of change in contrast and homogeneity: water and urea yielded smoother deposits with greater homogeneity, calcium and starch, rougher deposits with less homogeneity. Within each family, adulterant identity was further resolved urea by its extreme homogeneity increase and entropy collapse at higher concentration, and calcium versus starch by divergent AUC trends and granular percentage magnitude. (1) Both water and urea adulteration correlated with decreasing contrast (Water, ST: 0.220 – 0.129 vs. DT: 0.169 – 0.121), (Urea, ST: 0.223 – 0.075 vs. DT: 0.223 – 0.134) and increasing homogeneity (Water, ST: 0.924 – 0.970 vs. DT: 0.0.959 – 0.979), (Urea, ST: 0.923 – 0.991 vs. DT: 0.927 – 0.998) relative to baseline. (2) Both starch and calcium adulteration correlated with increasing contrast (Starch, ST: 0.184 – 0.397 vs. DT: 0.214 – 0.411), (Calcium, ST: 0.307 – 0.495 vs. DT: 0.234 – 0.446) and decreasing homogeneity (Starch, ST: 0.942 – 0.882 vs. DT: 0.929 – 0.876), (Calcium, ST: 0.908 – 0.849 vs. DT: 0.0.938 – 0.878) relative to baseline. (3) Among the evaluated image-derived parameters, granular percentage showed consistent discriminatory potential across water, urea, calcium, and starch adulteration. Its values varied distinctly between ST and DT milk, with ranges of 1.05–0.26% vs. 0.09–0.05% for water, 0.98–0% vs. 0.77–0.03% for urea, 0.55–0.88% vs. 0.21–0.51% for calcium, and 0.74–6.93% vs. 0.97–2.99% for starch, respectively.

These results support the central premise of this work: that microstructural signatures captured through simple optical microscopy and image analysis carry sufficient information to screen for milk adulteration without the need for chemical reagents, specialized instrumentation, or trained laboratory personnel.

Nevertheless, although a robust hierarchical classification framework has been proposed for adulterant identification, the current study is based on a limited number of dried droplet samples for each adulterant category. Future work should therefore be directed toward a full statistical characterization of a bigger dataset with multiple replicate droplets for each individual adulterant to increase the statistical power and generalizability of the classification outcomes. Moreover, the present qualitative/semi-quantitative classification approach should be extended to full end-to-end machine learning framework that can accurately quantify the concentration of adulterants through integrated image analysis and model prediction rather than categorical classification alone.

**Data availability:** The data pertaining to this work is present in the main article and supporting information documents.

**Competing interests:** The authors do not have any conflicts of interest.

**Acknowledgements:** NG thanks IIT Kharagpur for the Post-doctoral Research Fellowship. PD thanks ANRF for partially funding this research (vide the ARG grant). The authors thank IIT Kharagpur for partially funding this research through internal grants.

## References

[1] P.C. Pereira, Milk nutritional composition and its role in human health, Nutrition 30 (2014) 619–627. https://doi.org/10.1016/j.nut.2013.10.011.

[2] P.J. Huth, D.B. DiRienzo, G.D. Miller, Major Scientific Advances with Dairy Foods in Nutrition and Health, J. Dairy Sci. 89 (2006) 1207–1221. https://doi.org/10.3168/jds.S0022-0302(06)72190-7.

[3] M.K. Nieuwoudt, S.E. Holroyd, C.M. McGoverin, M.C. Simpson, D.E. Williams, Rapid, sensitive, and reproducible screening of liquid milk for adulterants using a portable Raman spectrometer and a simple, optimized sample well, J. Dairy Sci. 99 (2016) 7821–7831. https://doi.org/10.3168/jds.2016-11100.

[4] M.F. Mabrook, M.C. Petty, A novel technique for the detection of added water to full fat milk using single frequency admittance measurements, Sensors Actuators B Chem. 96 (2003) 215–218. https://doi.org/10.1016/S0925-4005(03)00527-6.

[5] H. Singuluri, Milk Adulteration in Hyderabad, India – A Comparative Study on the Levels of Different Adulterants Present in Milk, J. Chromatogr. Sep. Tech. 05 (2014). https://doi.org/10.4172/2157-7064.1000212.

[6] A.S. Dickerson, J.S. Lee, C. Keshava, A. Hotchkiss, A.S. Persad, Assessment of Health Effects of Exogenous Urea: Summary and Key Findings, Curr. Environ. Heal. Reports 5 (2018) 205–212. https://doi.org/10.1007/s40572-018-0198-8.

[7] M.K. Picolos, P.R. Orlander, Calcium Carbonate Toxicity: The Updated Milk-Alkali Syndrome; Report of 3 Cases and Review of the Literature, Endocr. Pract. 11 (2005) 272–280. https://doi.org/10.4158/EP.11.4.272.

[8] T. Azad, S. Ahmed, Common milk adulteration and their detection techniques, Int. J. Food Contam. 3 (2016) 22. https://doi.org/10.1186/s40550-016-0045-3.

[9] S. Das, B. Goswami, K. Biswas, Milk Adulteration and Detection: A Review, Sens. Lett. 14 (2016) 4–18. https://doi.org/10.1166/sl.2016.3580.

[10] A. Poonia, A. Jha, R. Sharma, H.B. Singh, A.K. Rai, N. Sharma, Detection of adulteration in milk: A review, Int. J. Dairy Technol. 70 (2017) 23–42. https://doi.org/10.1111/1471-0307.12274.

[11] G. Abernethy, K. Higgs, Rapid detection of economic adulterants in fresh milk by liquid chromatography–tandem mass spectrometry, J. Chromatogr. A 1288 (2013) 10–20. https://doi.org/10.1016/j.chroma.2013.02.022.

[12] M.K. Nieuwoudt, S.E. Holroyd, C.M. McGoverin, M.C. Simpson, D.E. Williams, Raman spectroscopy as an effective screening method for detecting adulteration of milk with small

nitrogen-rich molecules and sucrose, J. Dairy Sci. 99 (2016) 2520–2536. https://doi.org/10.3168/jds.2015-10342.

[13] Y. Cheng, Y. Dong, J. Wu, X. Yang, H. Bai, H. Zheng, D. Ren, Y. Zou, M. Li, Screening melamine adulterant in milk powder with laser Raman spectrometry, J. Food Compos. Anal. 23 (2010) 199–202. https://doi.org/10.1016/j.jfca.2009.08.006.

[14] P.M. Santos, E.R. Pereira-Filho, L.A. Colnago, Detection and quantification of milk adulteration using time domain nuclear magnetic resonance (TD-NMR), Microchem. J. 124 (2016) 15–19. https://doi.org/10.1016/j.microc.2015.07.013.

[15] O. Cirak, N.C. Icyer, M.Z. Durak, Rapid detection of adulteration of milks from different species using Fourier Transform Infrared Spectroscopy (FTIR), J. Dairy Res. 85 (2018) 222–225. https://doi.org/10.1017/S0022029918000201.

[16] C.F. Nascimento, P.M. Santos, E.R. Pereira-Filho, F.R.P. Rocha, Recent advances on determination of milk adulterants, Food Chem. 221 (2017) 1232–1244. https://doi.org/10.1016/j.foodchem.2016.11.034.

[17] R. Sangubotla, A. Mastan, J. Kim, Recent Advances in Electrochemical Biosensors for the Detection of Milk Adulterants, Biosensors 16 (2026) 92. https://doi.org/10.3390/bios16020092.

[18] S. Tripathy, A.R. Ghole, K. Deep, S.R.K. Vanjari, S.G. Singh, A comprehensive approach for milk adulteration detection using inherent bio-physical properties as 'Universal Markers': Towards a miniaturized adulteration detection platform, Food Chem. 217 (2017) 756–765. https://doi.org/10.1016/j.foodchem.2016.09.037.

[19] V. Ruiz-Valdepeñas Montiel, E. Povedano, S. Benedé, L. Mata, P. Galán-Malo, M. Gamella, A.J. Reviejo, S. Campuzano, J.M. Pingarrón, Disposable Amperometric Immunosensor for the Detection of Adulteration in Milk through Single or Multiplexed Determination of Bovine, Ovine, or Caprine Immunoglobulins G, Anal. Chem. 91 (2019) 11266–11274. https://doi.org/10.1021/acs.analchem.9b02336.

[20] N. Seddaoui, R. Attaallah, A. Amine, Development of an optical immunoassay based on peroxidase-mimicking Prussian blue nanoparticles and a label-free electrochemical immunosensor for accurate and sensitive quantification of milk species adulteration, Microchim. Acta 189 (2022) 209. https://doi.org/10.1007/s00604-022-05302-9.

[21] T.M. Samuel, Q. Zhou, F. Giuffrida, D. Munblit, V. Verhasselt, S.K. Thakkar, Nutritional and Non-nutritional Composition of Human Milk Is Modulated by Maternal, Infant, and Methodological Factors, Front. Nutr. 7 (2020). https://doi.org/10.3389/fnut.2020.576133.

[22] N. Shahidzadeh, M.F.L. Schut, J. Desarnaud, M. Prat, D. Bonn, Salt stains from evaporating droplets, Sci. Rep. 5 (2015) 10335. https://doi.org/10.1038/srep10335.

[23] S.A. McBride, S. Dash, S. Khan, K.K. Varanasi, Evaporative Crystallization of Spirals, Langmuir 35 (2019) 10484–10490. https://doi.org/10.1021/acs.langmuir.9b01002.

[24] J. Park, J. Moon, Control of Colloidal Particle Deposit Patterns within Picoliter Droplets Ejected by Ink-Jet Printing, Langmuir 22 (2006) 3506–3513. https://doi.org/10.1021/la053450j.

[25] R.K. Dewan, V.A. Bloomfield, A. Chudgar, C.V. Morr, Viscosity and Voluminosity of Bovine Milk Casein Micelles, J. Dairy Sci. 56 (1973) 699–705. https://doi.org/10.3168/jds.S0022-0302(73)85236-1.

[26] V. Kumar, S. Dash, Evaporation-Based Low-Cost Method for the Detection of Adulterant in Milk, ACS Omega 6 (2021) 27200–27207. https://doi.org/10.1021/acsomega.1c03887.

[27] C. Guillaume, S. Marchesseau, A. Lagaude, J.-L. Cuq, Effect of Salt Addition on the Micellar Composition of Milk Subjected to pH Reversible CO2 Acidification, J. Dairy Sci. 85 (2002) 2098–2105. https://doi.org/10.3168/jds.S0022-0302(02)74287-2.

[28] K.M. Brown, W.R. McManus, D.J. McMahon, Starch addition in renneted milk gels: Partitioning between curd and whey and effect on curd syneresis and gel microstructure, J. Dairy Sci. 95 (2012) 6871–6881. https://doi.org/10.3168/jds.2011-5191.

[29] P. Ishwarya S, V.R. Dugyala, S. Pradhan, M.G. Basavaraj, Sessile drop evaporation approach to detect starch adulteration in milk, Food Control 143 (2023) 109272. https://doi.org/10.1016/j.foodcont.2022.109272.

[30] N. Fu, M.W. Woo, X.D. Chen, Colloidal transport phenomena of milk components during convective droplet drying, Colloids Surfaces B Biointerfaces 87 (2011) 255–266. https://doi.org/10.1016/j.colsurfb.2011.05.026.

[31] E.M. Both, M. Nuzzo, A. Millqvist-Fureby, R.M. Boom, M.A.I. Schutyser, Morphology development during single droplet drying of mixed component formulations and milk, Food Res. Int. 109 (2018) 448–454. https://doi.org/10.1016/j.foodres.2018.04.043.

[32] C. Sadek, L. Pauchard, P. Schuck, Y. Fallourd, N. Pradeau, C. Le Floch-Fouéré, R. Jeantet, Mechanical properties of milk protein skin layers after drying: Understanding the mechanisms of particle formation from whey protein isolate and native phosphocaseinate, Food Hydrocoll. 48 (2015) 8–16. https://doi.org/10.1016/j.foodhyd.2015.01.014.

[33] B. Pathak, J. Christy, Evaporation dynamics of a sessile milk droplet placed on a hydrophobic surface, Colloids Surfaces A Physicochem. Eng. Asp. 665 (2023) 131207. https://doi.org/10.1016/j.colsurfa.2023.131207.

[34] S. Patari, P. Datta, P.S. Mahapatra, 3D Paper-based milk adulteration detection device, Sci. Rep. 12 (2022) 13657. https://doi.org/10.1038/s41598-022-17851-3.

[35] A. Mamgain, V. Kumar, S. Dash, Image-Based Detection of Adulterants in Milk Using Convolutional Neural Network, ACS Omega 9 (2024) 27158–27168. https://doi.org/10.1021/acsomega.4c01274.

[36] J.S. Farah, R.N. Cavalcanti, J.T. Guimarães, C.F. Balthazar, P.T. Coimbra, T.C. Pimentel, E.A. Esmerino, M.C.K.H. Duarte, M.Q. Freitas, D. Granato, R.P.C. Neto, M.I.B. Tavares, V. Calado, M.C. Silva, A.G. Cruz, Differential scanning calorimetry coupled with machine learning technique: An effective approach to determine the milk authenticity, Food Control 121 (2021) 107585. https://doi.org/10.1016/j.foodcont.2020.107585.

[37] C. Chu, H. Wang, X. Luo, Y. Fan, L. Nan, C. Du, D. Gao, P. Wen, D. Wang, Z. Yang, G. Yang, L. Liu, Y. Li, B. Hu, A. Zunongjiang, S. Zhang, Rapid detection and quantification of melamine, urea, sucrose, water, and milk powder adulteration in pasteurized milk using Fourier transform infrared (FTIR) spectroscopy coupled with modern statistical machine learning algorithms, Heliyon 10 (2024) e32720. https://doi.org/10.1016/j.heliyon.2024.e32720.

[38] A. Sharma, M. Yadav, N. Kumar, G. Chandu, R. Jana, P. Pandey, R. Kumar, Rapid detection of milk adulteration using AI-driven portable colorimetric spectroscopy, Discov. Food 6 (2026) 118. https://doi.org/10.1007/s44187-026-00855-7.

[39] A. Sitorus, R. Lapcharoensuk, Exploring Deep Learning to Predict Coconut Milk Adulteration Using FT-NIR and Micro-NIR Spectroscopy, Sensors 24 (2024) 2362. https://doi.org/10.3390/s24072362.

[40] M. Kumar, R. Kumar, A. Chakravorty, A Low-Cost Image Histogram and Machine Learning Approach for Detection of Cow Milk Adulteration, J. Food Sci. 91 (2026). https://doi.org/10.1111/1750-3841.70831.

[41] H.A. Neto, W.L.F. Tavares, D.C.S.Z. Ribeiro, R.C.O. Alves, L.M. Fonseca, S.V.A. Campos, On the utilization of deep and ensemble learning to detect milk adulteration, BioData Min. 12 (2019) 13. https://doi.org/10.1186/s13040-019-0200-5.

[42] R.D. Deegan, O. Bakajin, T.F. Dupont, G. Huber, S.R. Nagel, T.A. Witten, Capillary flow as the cause of ring stains from dried liquid drops, Nature 389 (1997) 827–829. https://doi.org/10.1038/39827.

[43] E. Boel, R. Koekoekx, S. Dedroog, I. Babkin, M.R. Vetrano, C. Clasen, G. Van den Mooter, Unraveling Particle Formation: From Single Droplet Drying to Spray Drying and Electrospraying, Pharmaceutics 12 (2020) 625. https://doi.org/10.3390/pharmaceutics12070625.

[44] J.P. Barger, P.F. Dillon, Near-membrane electric field calcium ion dehydration, Cell Calcium 60 (2016) 415–422. https://doi.org/10.1016/j.ceca.2016.09.006.

[45] P. Hristov, I. Mitkov, D. Sirakova, I. Mehandgiiski, G. Radoslavov, Measurement of Casein Micelle Size in Raw Dairy Cattle Milk by Dynamic Light Scattering, in: Milk Proteins - From Struct. to Biol. Prop. Heal. Asp., InTech, 2016. https://doi.org/10.5772/62779.

[46] C. Fuentes, I. Kang, J. Lee, D. Song, M. Sjöö, J. Choi, S. Lee, L. Nilsson, Fractionation and characterization of starch granules using field-flow fractionation (FFF) and differential scanning calorimetry (DSC), Anal. Bioanal. Chem. 411 (2019) 3665–3674. https://doi.org/10.1007/s00216-019-01852-9.

[47] L. Hua, R. Zhou, D. Thirumalai, B.J. Berne, Urea denaturation by stronger dispersion interactions with proteins than water implies a 2-stage unfolding, Proc. Natl. Acad. Sci. 105 (2008) 16928–16933. https://doi.org/10.1073/pnas.0808427105.

[48] DLVO theory of colloid stability, in: 2024: pp. 217–244. https://doi.org/10.1016/B978-0-443-16116-2.00009-6.

# Supporting information document

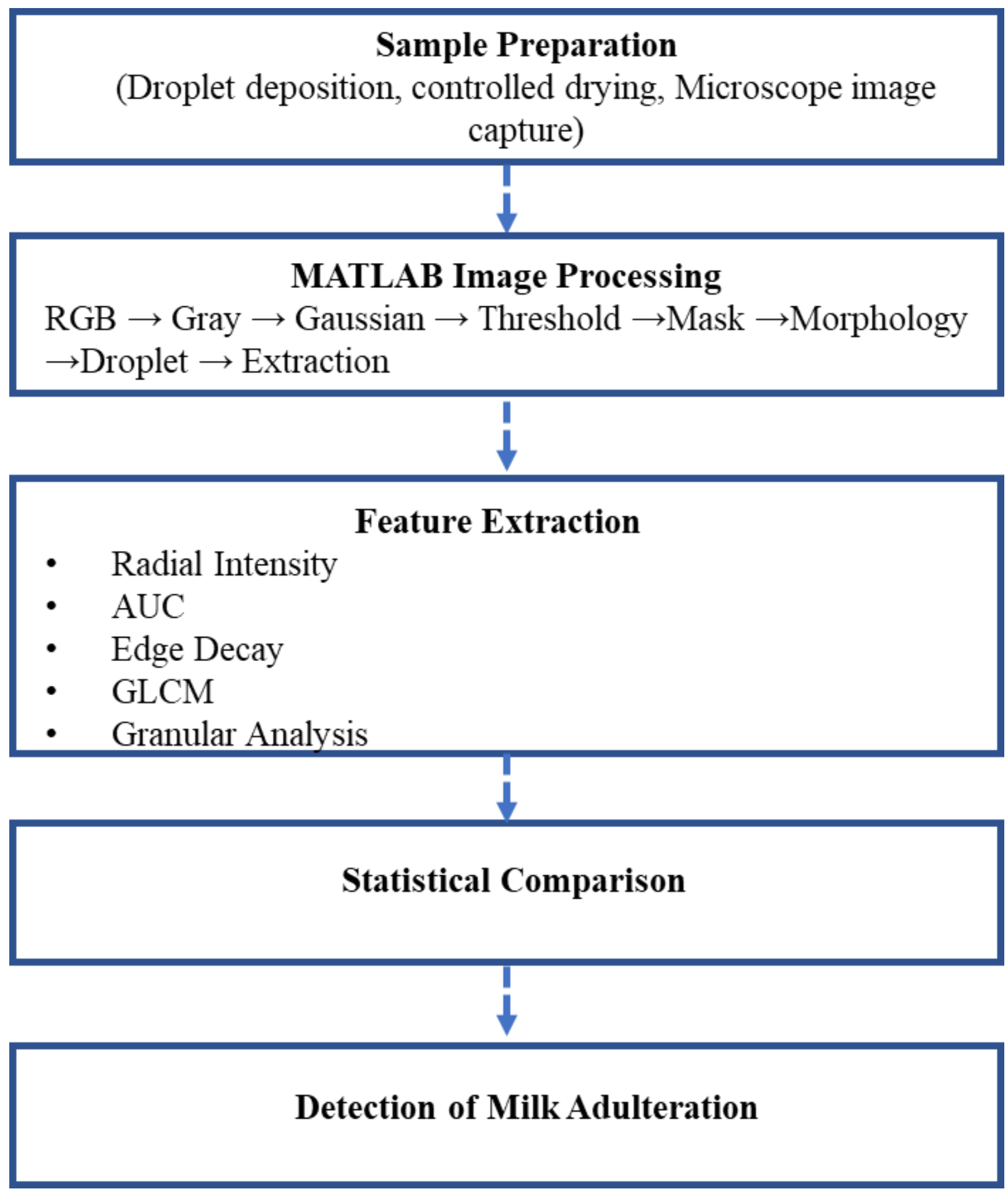

**Fig.S1** Schematic workflow illustrating the sequential steps involved in sample preparation, drying-pattern generation, microscopic image acquisition, image processing, and quantitative analysis.

Single Toned Milk

200 µm 200 µm 200 µm 200 µm 200 µm 200 µm

0% Water 10% Water 20% Water 30% Water 40% Water 50% Water

Double Toned Milk

200 µm 200 µm 200 µm 200 µm 200 µm 200 µm

0% Water 10% Water 20% Water 30% Water 40% Water 50% Water

**Fig. S2** Representative microscopic images illustrating the evolution of drying patterns of milk droplets at different levels of water dilution. The images demonstrate progressive changes in radial structures, peripheral deposition, and internal drying morphology with increasing dilution (Scale bar 200 µm).

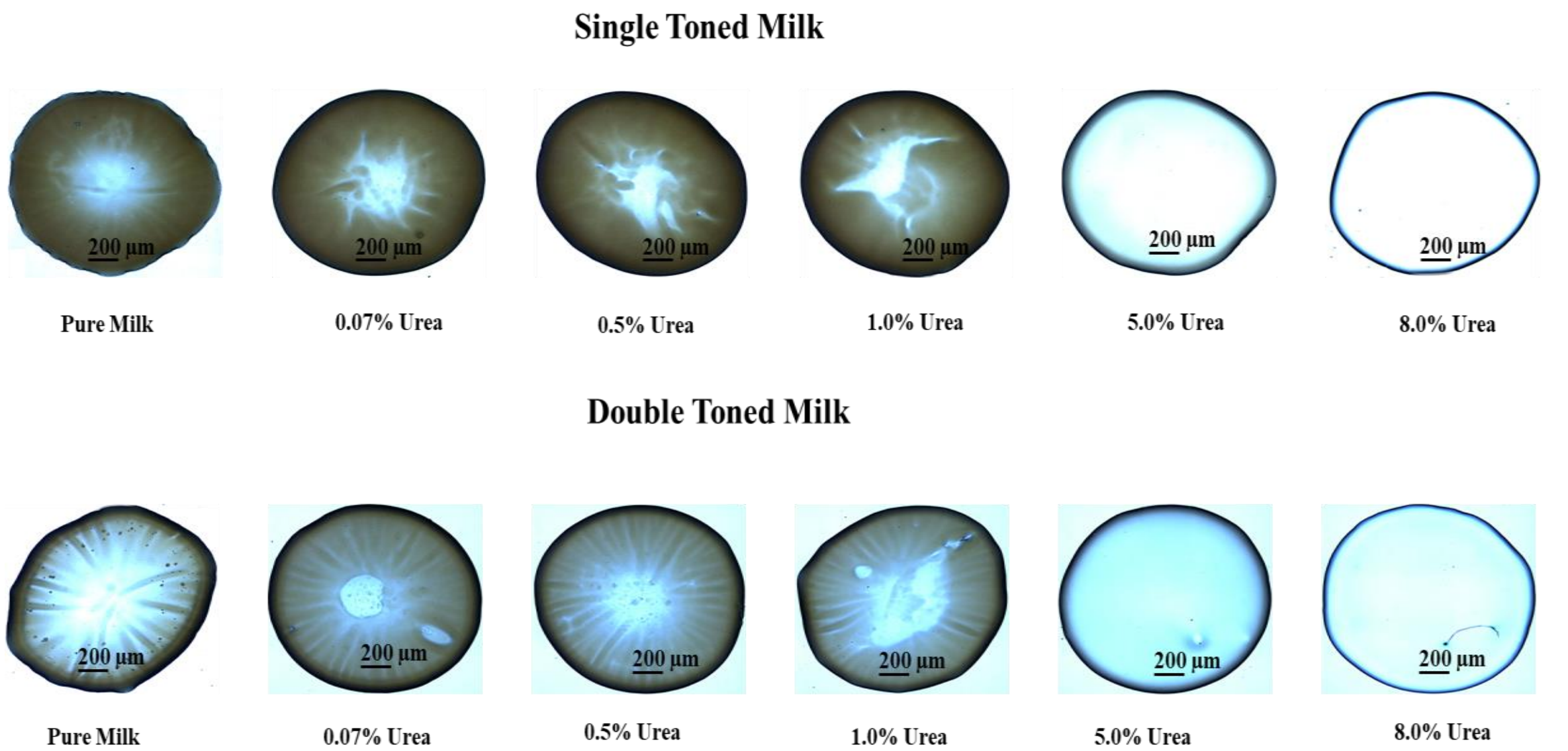

**Fig. S3** Representative microscopic images illustrating the evolution of drying patterns of milk droplets at different levels of urea adulteration. The images demonstrate progressive changes in radial structures, peripheral deposition, and internal drying morphology with increasing urea adulteration (Scale bar 200 µm).

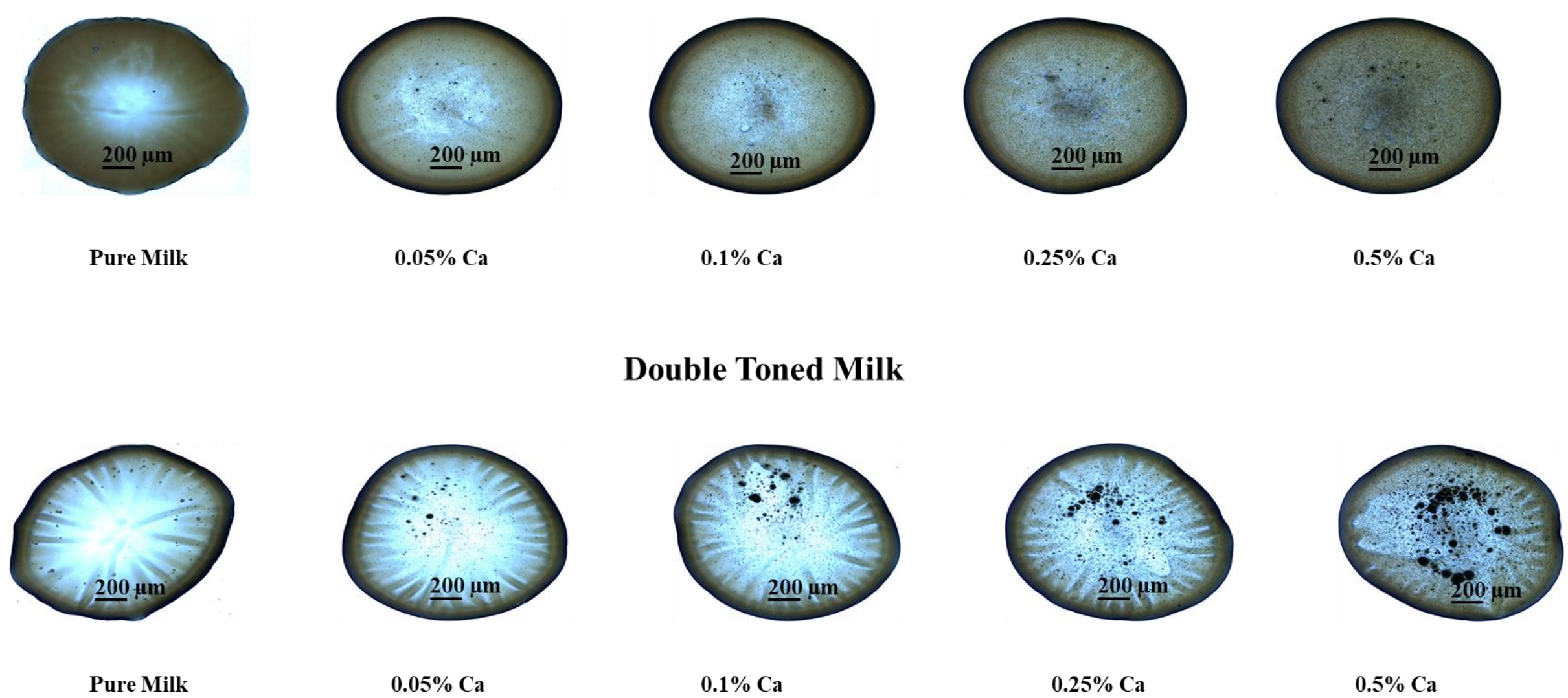

**Fig. S4** Representative microscopic images illustrating the evolution of drying patterns of milk droplets at different levels of calcium adulteration. The images demonstrate progressive changes in radial structures, peripheral deposition, and internal drying morphology with increasing calcium adulteration (Scale bar 200 µm).

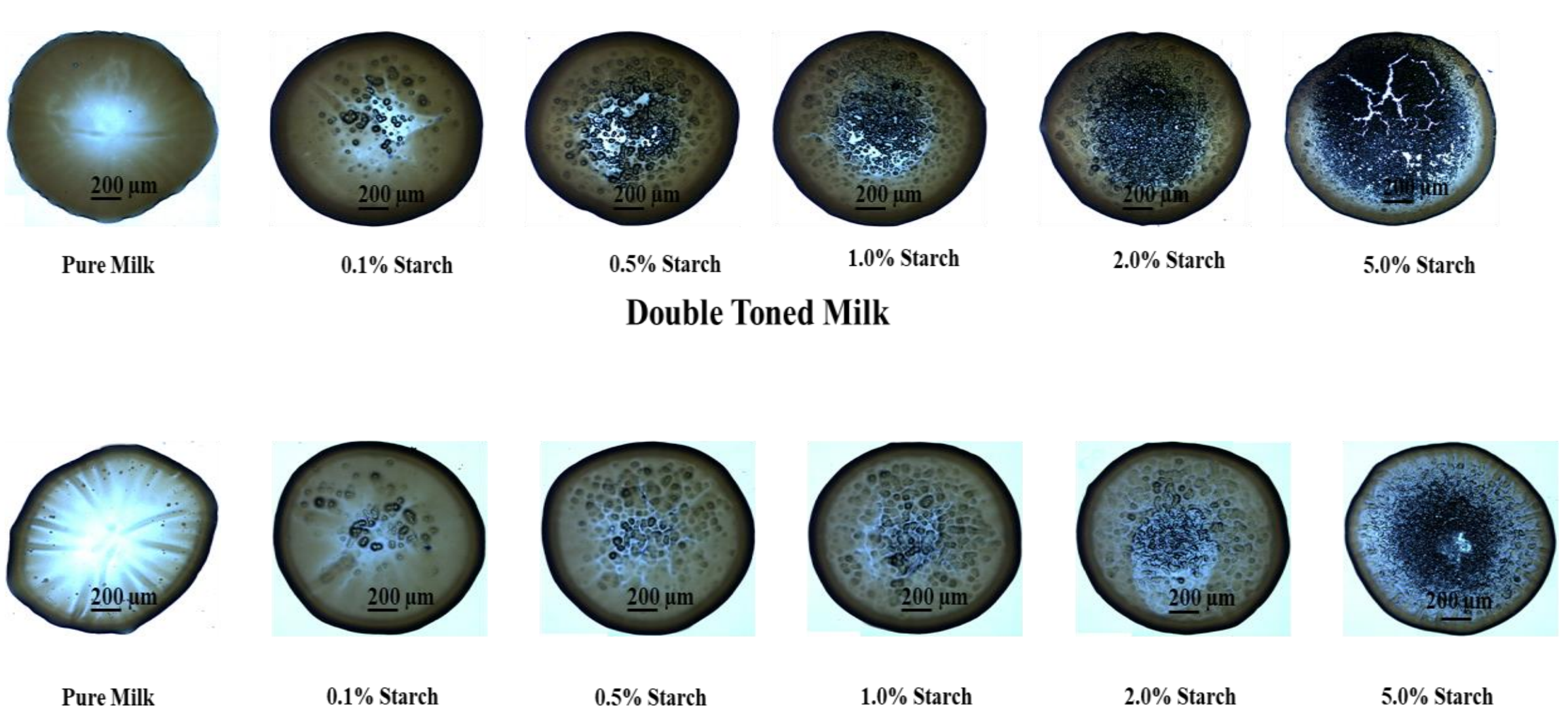

**Fig. S5** Representative microscopic images illustrating the evolution of drying patterns of milk droplets at different levels of starch adulteration. The images demonstrate progressive changes in radial structures, peripheral deposition, and internal drying morphology with increasing starch adulteration (Scale bar 200 µm).

**Table S1:** ST milk data

| S.N | Adulterant | % | Contrast | Homogeneity | Granular % |
|---|---|---|---|---|---|
| 1 | Water | 0 | 0.2197 | 0.9239 | 1.05 |
| 2 | Water | 10 | 0.1744 | 0.9444 | 0.81 |
| 3 | Water | 20 | 0.182 | 0.9468 | 0.42 |
| 4 | Water | 30 | 0.1707 | 0.9511 | 0.37 |
| 5 | Water | 40 | 0.1474 | 0.9637 | 0.13 |
| 6 | Water | 50 | 0.1293 | 0.9698 | 0.26 |
| 7 | Urea | 0.07 | 0.2234 | 0.9228 | 0.98 |
| 8 | Urea | 0.5 | 0.2021 | 0.9345 | 0.88 |
| 9 | Urea | 1 | 0.2222 | 0.9243 | 0.02 |
| 10 | Urea | 5 | 0.103 | 0.9793 | 0 |
| 11 | Urea | 8 | 0.0752 | 0.991 | 0 |
| 12 | Calcium | 0.05 | 0.3071 | 0.9077 | 0.55 |
| 13 | Calcium | 0.1 | 0.3676 | 0.891 | 0.5 |
| 14 | Calcium | 0.25 | 0.3819 | 0.8752 | 0.9 |
| 15 | Calcium | 0.5 | 0.4951 | 0.8494 | 0.88 |
| 16 | Starch | 0.1 | 0.184 | 0.9422 | 0.74 |
| 17 | Starch | 0.5 | 0.2032 | 0.9383 | 1.65 |
| 18 | Starch | 1 | 0.24 | 0.9249 | 1.86 |
| 19 | Starch | 2 | 0.3363 | 0.8996 | 4.12 |
| 20 | Starch | 5 | 0.3966 | 0.8821 | 6.93 |

**Table S2:** DT milk data

| S.N | Adulterant | % | Contrast | Homogeneity | Granular % |
|---|---|---|---|---|---|
| 1 | Water | 0 | 0.1692 | 0.9593 | 0.09 |
| 2 | Water | 10 | 0.1412 | 0.9701 | 0.06 |
| 3 | Water | 20 | 0.1339 | 0.9723 | 0.05 |
| 4 | Water | 30 | 0.1378 | 0.9727 | 0.05 |
| 5 | Water | 40 | 0.1306 | 0.9751 | 0.02 |
| 6 | Water | 50 | 0.1211 | 0.9791 | 0.05 |
| 7 | Urea | 0.07 | 0.2228 | 0.9268 | 0.77 |
| 8 | Urea | 0.5 | 0.1978 | 0.9385 | 0.6 |
| 9 | Urea | 1 | 0.2074 | 0.933 | 0.56 |
| 10 | Urea | 5 | 0.1581 | 0.9531 | 0.04 |
| 11 | Urea | 8 | 0.0752 | 0.9976 | 0.03 |
| 12 | Calcium | 0.05 | 0.2342 | 0.938 | 0.21 |
| 13 | Calcium | 0.1 | 0.2333 | 0.9308 | 0.19 |
| 14 | Calcium | 0.25 | 0.3069 | 0.9094 | 0.33 |
| 15 | Calcium | 0.5 | 0.4457 | 0.8776 | 0.51 |
| 16 | Starch | 0.1 | 0.2143 | 0.9294 | 0.97 |
| 17 | Starch | 0.5 | 0.2284 | 0.9241 | 0.64 |
| 18 | Starch | 1 | 0.3164 | 0.8923 | 1.36 |
| 19 | Starch | 2 | 0.5025 | 0.8487 | 2.95 |
| 20 | Starch | 5 | 0.4111 | 0.8762 | 2.99 |

## S1. Statistical characterisation of descriptor responses

### S1.1 Dose-response behaviour

The rank-correlation and regression analysis for each series of adulterant is presented in Table S3. Of the four adulterants, water dilution produced the most consistent monotonic behavior. Homogeneity in toned milk increased in a perfectly monotonic manner with dilution ($\rho = 1.000$, $p < 0.001$, $R^2 = 0.932$, $p = 0.002$), while both contrast and granular area decreased ($\rho = -0.943$, $p = 0.005$, for both). The same pattern was observed in double-toned milk (homogeneity $\rho = 1.000$; contrast $\rho = -0.943$, $p = 0.005$) confirming that dilution is a smooth, non-saturating rescaling of deposit structure without change in its qualitative character.

Of all the adulterants, calcium gave the most regular and strongest contrast response. In toned milk, the rank correlation with concentration was perfect for both contrast ($\rho = 1.000$) and homogeneity ($\rho = -1.000$), with regression on log concentration explaining 94.9% and 96.8% of variance respectively ($p = 0.005$ and $p = 0.002$). Reproduction of this behavior was made in double-toned milk (contrast $\rho = 0.900$, $p = 0.037$; homogeneity $\rho = -1.000$, R 2 = 0.947, $p = 0.005$). Granular area was independent of calcium in toned milk ($\rho = -0.200$, $p = 0.747$, R2 = 0.005) but highly correlated with calcium in double-toned milk ($\rho = 0.900$, R2 = 0.887, $p = 0.017$), suggesting that calcium acts through colloidal aggregation of the pre-existing protein phase rather than through the generation of a distinct particulate population, so that the capacity to detect calcium in the granular channel is dependent on the background particulate loading of the milk matrix.

Greatest granular response was produced by starch as an insoluble particulate adulterant. The granular area increased sharply and monotonically for both types of milk ($\rho = 0.943$, $p = 0.005$), with regression slopes of 2.74 and 1.46 per $\log_{10}$ unit concentration in ST and DT respectively – the largest coefficients ever recorded for any adulterant-descriptor combination in the study.

Urea exhibited the most divergent behavior between milk types. The granular area showed the strongest monotonic trend of the whole dataset in toned milk ($\rho = -0.986$, $p < 0.001$), where the granular area decreased to zero at 5% and 8% urea, and homogeneity increased significantly ($\rho = 0.886$, $p = 0.019$). No significant differences were obtained for any of the three descriptors for the whole series of double toned milk (contrast $\rho = -0.600$, $p = 0.208$; homogeneity $\rho = 0.371$, $p = 0.469$; granular $\rho = -0.657$, $p = 0.156$). Inspection of the underlying values shows that this is a truly non-monotonic, threshold-type response rather than an absence of effect: DT descriptors change relatively little between 0.07% and 5% urea before shifting abruptly at 8% (homogeneity 0.9531 to 0.9976; contrast 0.1581 to 0.0752). Rank-based trend statistics that assume monotonicity are blind to precisely this shape of response and the failure to reach significance should thus be interpreted as evidence for a threshold rather than for a null effect.

**S1.2 Comparison of toned milk and double toned milk**

Granular area was significantly higher in toned milk than double-toned milk for the 20 matched adulterant–concentration conditions (mean 1.153% vs 0.624%; Wilcoxon $W = 23.0$, $p = 0.002$; paired $t = 2.62$, $p = 0.017$; Cohen's d z = 0.586). For contrast and homogeneity, no significant differences were found between milk types (contrast $p = 0.295$; homogeneity $p = 0.179$) and both had small effect sizes ($|d_z| \leq 0.29$). It implies that the milk matrix affects the particulate channel of the analysis to a much greater extent than it affects the textural channels and that discrimination based on contrast and homogeneity is relatively insensitive to milk type – a practically useful outcome for any screening application that is deployed, as it means these two descriptors do not have to be re-calibrated separately for each milk grade.

**S1.3 Intercorrelation between descriptors and dimensionality**

Contrast and homogeneity were strongly and inversely correlated in the pooled dataset (Pearson $r = -0.979$, $p = 6.0 \times 10^{-28}$; Spearman $\rho = -0.978$) confirming that the two GLCM descriptors are largely redundant and capture a single underlying axis of deposit texture. Granular area was moderately correlated with both ($r = 0.564$ with contrast, $p = 1.5 \times 10^{-4}$; $r = -0.562$ with homogeneity, $p = 1.6 \times 10^{-4}$), suggesting it contains partially independent information.

This structure was confirmed by principal component analysis. PC1 explained 80.8% of total variance, with negative loading on homogeneity (−0.617) and positive loading on contrast (+0.617), representing the textural heterogeneity axis. PC2 accounted for an additional 18.5% and loaded mostly on granular area (+0.873) which may be interpreted as the particulate axis. Together the two components accounted for 99.3% of variance, thus the descriptor set is effectively two-dimensional: a texture axis and a particulate axis. This is a compact and interpretable result that maps directly onto the mechanistic distinction developed in **Sec. 3.1** between adulterants acting on the protein network and those introducing a discrete insoluble phase.

### S1.4 Multivariate discrimination of adulterants

A leave-one-out cross-validated classification of the four adulterant classes from the three descriptors with linear discriminant analysis yielded an accuracy of 65.0%, and 72.5% with a random forest, versus a 25% chance baseline. Looking at the LDA confusion matrix, calcium is the best-recovered class (6/8 correct). The main error source was confusion between urea and water (4 water samples incorrectly assigned to urea and 3 urea samples to water) and between starch and urea (4 starch samples misassigned). This is consistent with the dose-response results: low concentration urea and water both act primarily by reducing textural heterogeneity, and are thus truly similar in this three descriptor space at the low end of their ranges. The inclusion of the radial-profile descriptors encoding the spatial distribution of the deposit rather than its local texture and thus likely to separate exactly these classes, would be expected to improve discrimination.

### S1.5 Detection limits (indicative)

Table S4 shows regression-based detection limits. Calcium is the most sensitively detected adulterant in the contrast and homogeneity channels with estimated LOD values of 0.19–0.27% in toned milk and 0.10–0.14% in double-toned milk. The granular channel in toned milk was the best for starch detection (LOD ≈1.06%) which is consistent with its particulate mechanism. Urea resulted in the highest detection limits (2.5–2.6% in toned milk through contrast and homogeneity) which is a manifestation of the threshold nature of its response: the crystallisation-driven signal is only significant at high concentration, so the linear detection-limit model must necessarily perform poorly for this analyte and the reported values should be taken as conservative upper bounds rather than true sensitivity estimates. These numbers are given only as an indication, since they are based on single image measurements and a linear fitting model to few points.

### S1.6 Statistical constraints

The main limitation of this analysis is the absence of independent replication, with only one micrograph being taken per milk type per condition and, therefore, no estimate of between-droplet or between-session reproducibility, or inferential comparison of individual concentration levels. The trend, pairing and multivariate results presented above are valid within the limits of the concentration series, the ST/DT pairing and the pooled sample set, respectively, from which their degrees of freedom are calculated; they do not, however, prove that a given descriptor value would be reproduced on a repeat preparation of the same sample. The confidence intervals reported from the ROI are measures of the spatial heterogeneity within a deposit and should not be viewed as measures of the measurement uncertainty. Replicate droplets are required to create reproducibility and thus allow per-level hypothesis testing and formal method validation and are identified as the immediate priority for extending this work.

**Table S3** Dose–response statistics for each adulterant series. ρ = Spearman rank correlation with concentration; regression performed on $log_{10}$ concentration. Bold indicates $p < 0.05$.

| S.N. | Milk | Adulterant | Descriptor | n | ρ | p (ρ) | Slope | $R^2$ | p (reg) |
|---|---|---|---|---|---|---|---|---|---|
| 1 | ST | Water | Contrast | 6 | −0.943 | **0.005** | −0.075 | 0.863 | **0.007** |
| 2 | ST | Water | Homogeneity | 6 | 1.000 | **<0.001** | +0.041 | 0.932 | **0.002** |
| 3 | ST | Water | Granular | 6 | −0.943 | **0.005** | −0.880 | 0.925 | **0.002** |
| 4 | ST | Urea | Contrast | 6 | −0.714 | 0.111 | −0.061 | 0.710 | **0.035** |
| 5 | ST | Urea | Homogeneity | 6 | 0.886 | **0.019** | +0.029 | 0.729 | **0.030** |
| 6 | ST | Urea | Granular | 6 | −0.986 | **<0.001** | −0.515 | 0.804 | **0.015** |
| 7 | ST | Calcium | Contrast | 5 | 1.000 | **<0.001** | +0.193 | 0.949 | **0.005** |
| 8 | ST | Calcium | Homogeneity | 5 | −1.000 | **<0.001** | −0.056 | 0.968 | **0.002** |
| 9 | ST | Calcium | Granular | 5 | −0.200 | 0.747 | −0.033 | 0.005 | 0.911 |
| 10 | ST | Starch | Contrast | 6 | 0.829 | **0.042** | +0.093 | 0.666 | **0.048** |
| 11 | ST | Starch | Homogeneity | 6 | −0.657 | 0.156 | −0.023 | 0.538 | 0.097 |
| 12 | ST | Starch | Granular | 6 | 0.943 | **0.005** | +2.737 | 0.723 | **0.032** |
| 13 | DT | Water | Contrast | 6 | −0.943 | **0.005** | −0.041 | 0.919 | **0.003** |
| 14 | DT | Water | Homogeneity | 6 | 1.000 | **<0.001** | +0.017 | 0.953 | **0.001** |
| 15 | DT | Water | Granular | 6 | −0.820 | **0.046** | −0.051 | 0.744 | **0.027** |
| 16 | DT | Urea | Contrast | 6 | −0.600 | 0.208 | −0.035 | 0.373 | 0.197 |
| 17 | DT | Urea | Homogeneity | 6 | 0.371 | 0.469 | +0.014 | 0.245 | 0.319 |
| 18 | DT | Urea | Granular | 6 | −0.657 | 0.156 | −0.142 | 0.157 | 0.437 |
| 19 | DT | Calcium | Contrast | 5 | 0.900 | **0.037** | +0.192 | 0.859 | **0.023** |
| 20 | DT | Calcium | Homogeneity | 5 | −1.000 | **<0.001** | −0.059 | 0.947 | **0.005** |
| 21 | DT | Calcium | Granular | 5 | 0.900 | **0.037** | +0.297 | 0.887 | **0.017** |
| 22 | DT | Starch | Contrast | 6 | 0.943 | **0.005** | +0.150 | 0.745 | **0.027** |
| 23 | DT | Starch | Homogeneity | 6 | −0.943 | **0.005** | −0.049 | 0.813 | **0.014** |
| 24 | DT | Starch | Granular | 6 | 0.943 | **0.005** | +1.463 | 0.794 | **0.017** |

**Table S4.** Indicative regression-derived detection limits (LOD = 3s_res/|m|), fitted on native concentration scale including the 0% control.

| S.N. | **Milk** | **Adulterant** | **Descriptor** | **Slope (per %)** | **$R^2$** | **LOD (%)** |
|---|---|---|---|---|---|---|
| 1 | ST | Urea | Contrast | −0.0197 | 0.953 | 2.46 |
| 2 | ST | Urea | Homogeneity | +0.0091 | 0.948 | 2.59 |
| 3 | ST | Calcium | Contrast | +0.4680 | 0.869 | 0.27 |
| 4 | ST | Calcium | Homogeneity | −0.1381 | 0.934 | 0.19 |
| 5 | ST | Starch | Contrast | +0.0419 | 0.888 | 2.26 |
| 6 | ST | Starch | Granular | +1.2387 | 0.973 | 1.06 |
| 7 | DT | Urea | Contrast | −0.0144 | 0.805 | 5.47 |
| 8 | DT | Calcium | Contrast | +0.5187 | 0.981 | 0.10 |
| 9 | DT | Calcium | Homogeneity | −0.1508 | 0.963 | 0.14 |
| 10 | DT | Calcium | Granular | +0.7828 | 0.965 | 0.13 |

| 11 | DT | Starch | Granular | +0.5424 | 0.718 | 3.98 |
|---|---|---|---|---|---|---|

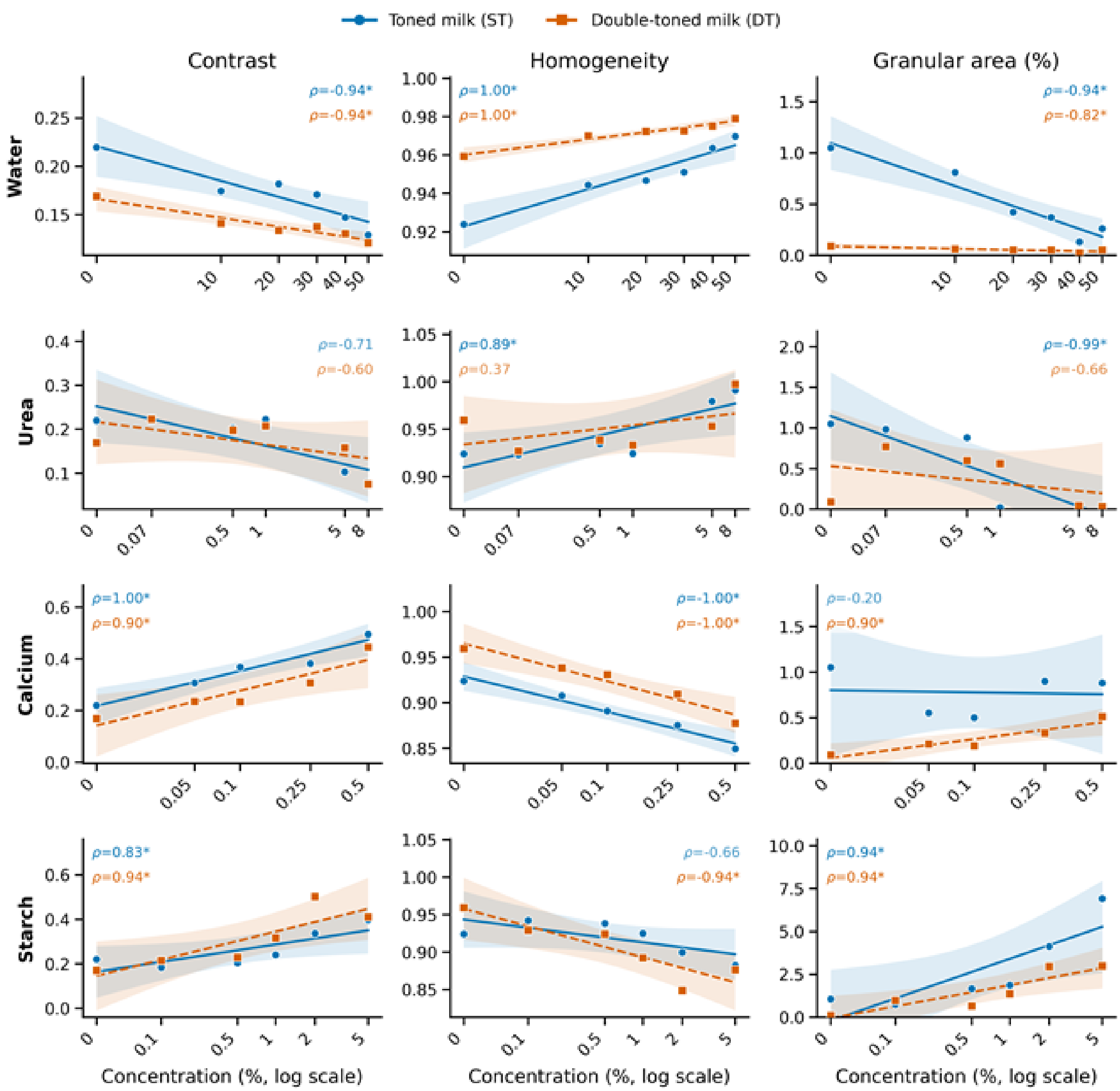

**Figure S6.** Dose–response behaviour of the three deposit descriptors for each adulterant. Rows correspond to adulterant (water, urea, calcium, starch) and columns to descriptor (contrast, homogeneity, granular area). Blue circles with solid lines denote toned milk (ST) and orange squares with dashed lines denote double-toned milk (DT). Lines are ordinary least-squares fits on $\log_{10}$ concentration with the 0% control included at an offset of half the lowest non-zero level; shaded regions are 95% confidence bands for the mean response. Spearman rank correlation coefficients are inset for each milk type; asterisks denote $p < 0.05$. The horizontal axis is logarithmic, with tick labels showing native concentration (%).

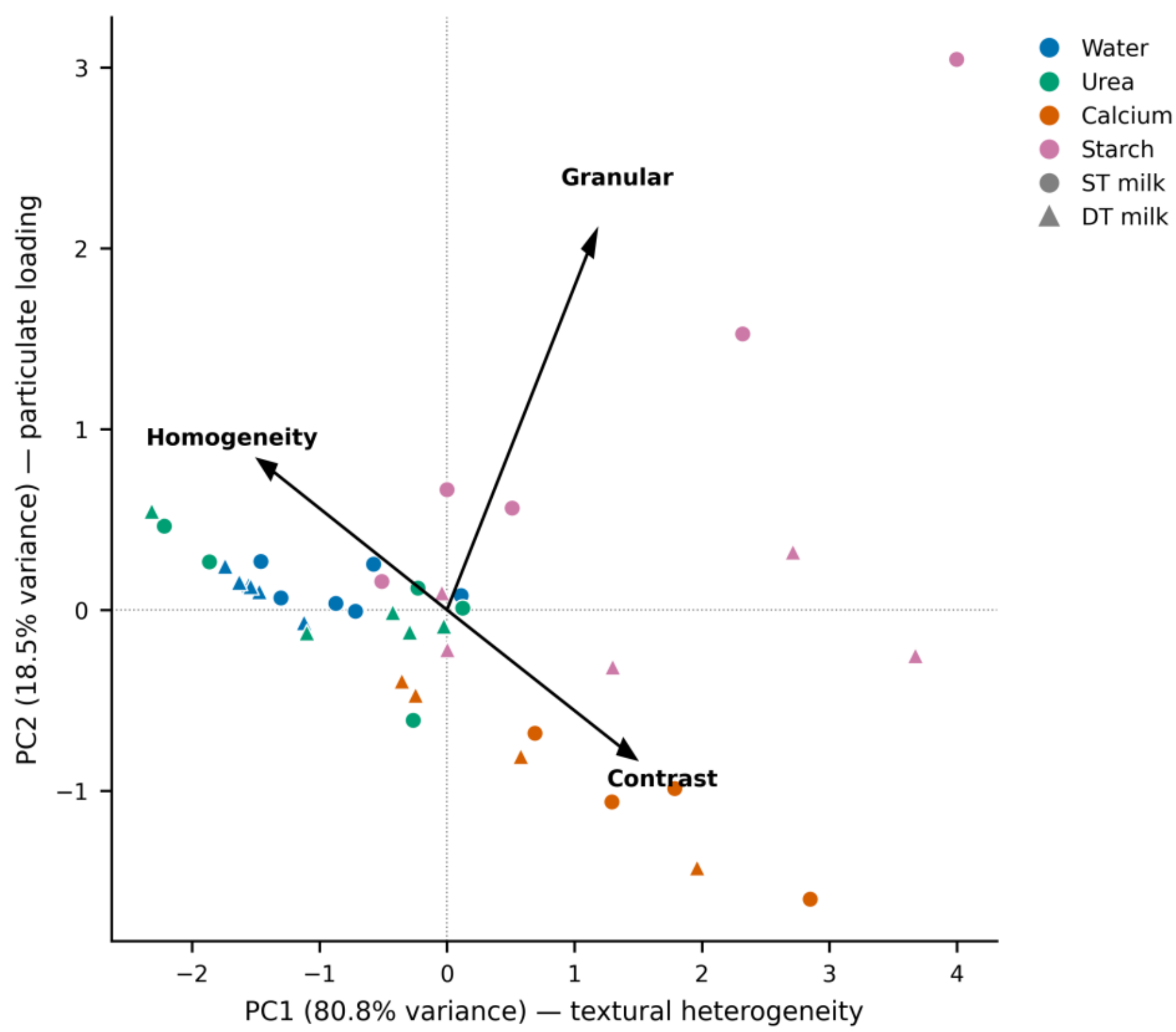

**Figure S7.** Principal component analysis of the standardised descriptor set across all 40 samples. Colour denotes adulterant and marker shape denotes milk type (circles, toned; triangles, double-toned). Black arrows are the descriptor loading vectors, scaled for display. PC1 (80.8% of variance) loads oppositely on contrast and homogeneity and represents textural heterogeneity; PC2 (18.5%) loads predominantly on granular area and represents particulate loading. The two components together account for 99.3% of total variance.

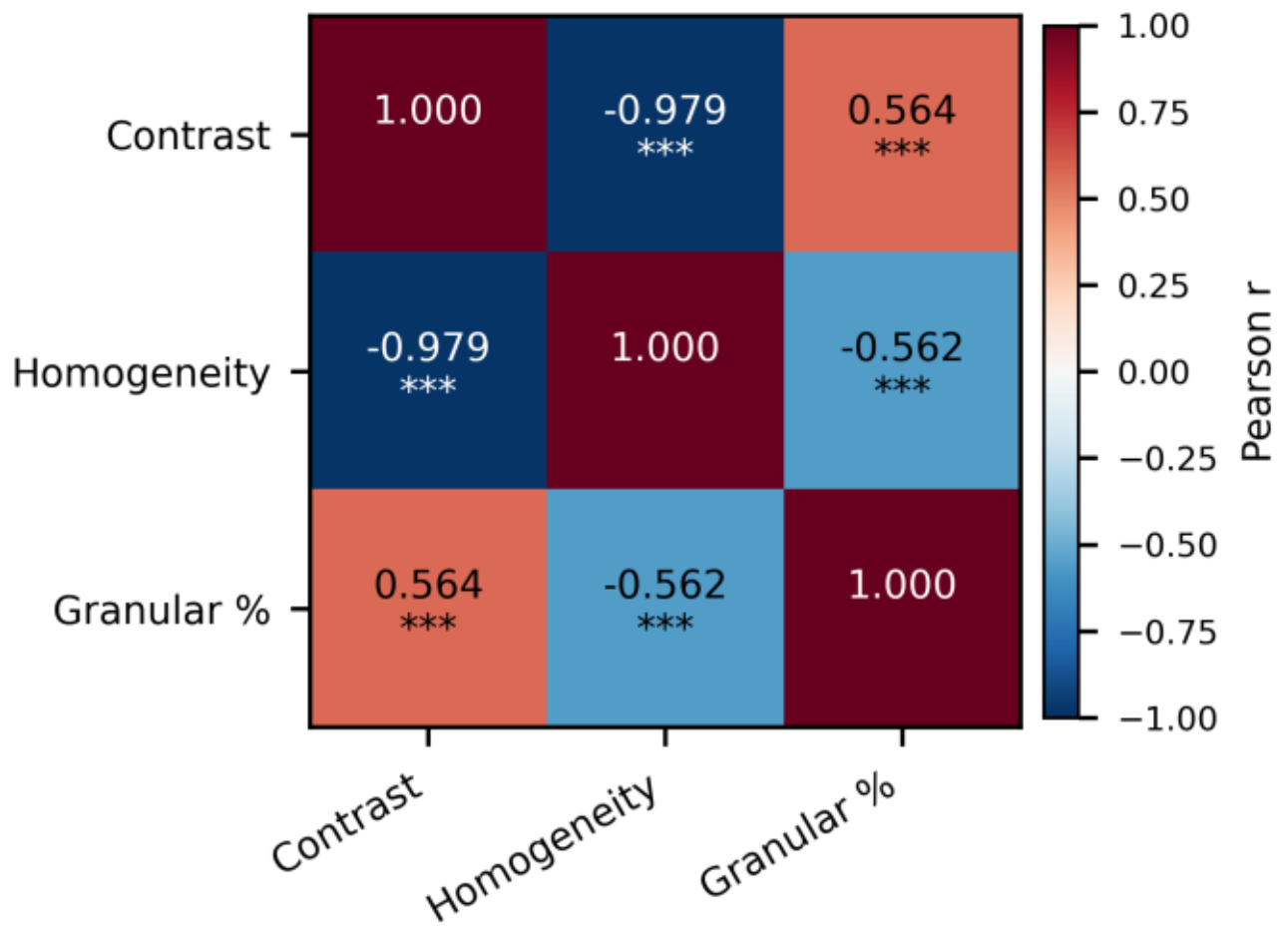

**Figure S8.** Pearson correlation matrix of the three deposit descriptors across the pooled dataset (n = 40). Asterisks denote significance: *p < 0.05, **p < 0.01, ***p < 0.001. The near-perfect inverse correlation between contrast and homogeneity (r = −0.979) indicates that these two GLCM descriptors capture a single underlying textural axis.

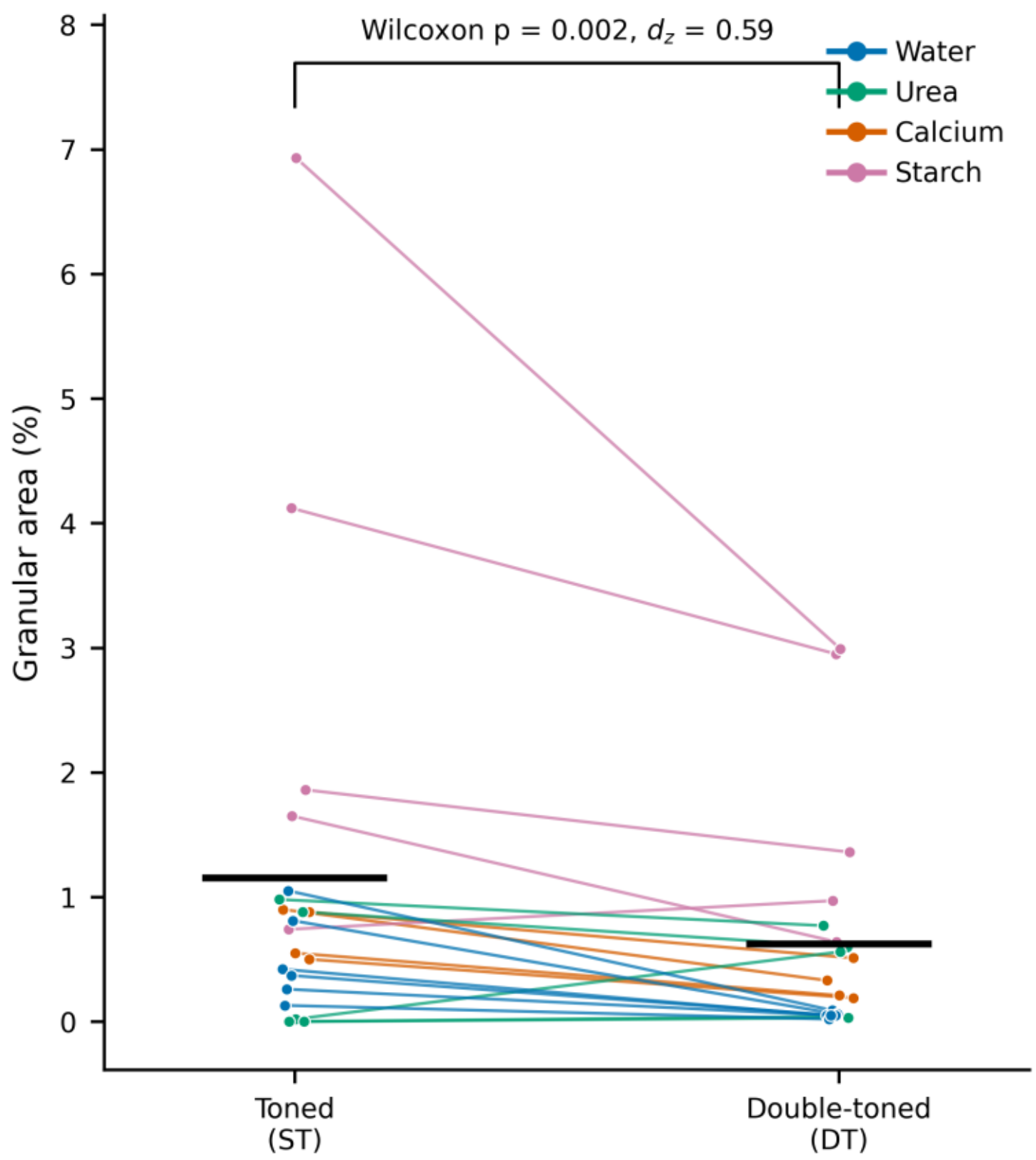

**Figure S9.** Paired comparison of granular area between toned and double-toned milk across the 20 matched adulterant–concentration conditions. Each line joins the two milk types for one condition, coloured by adulterant; horizontal black bars denote group means. Granular area was significantly higher in toned milk (Wilcoxon signed-rank p = 0.002; Cohen's d_z = 0.59). Points are horizontally jittered for visibility.